\documentclass[aps,prl,preprint,superscriptaddress,showpacs]{revtex4-1}

\usepackage{graphicx}

\begin{document}

\title{Solar irradiance variability is caused by the magnetic activity on the solar surface}

\author{Kok Leng Yeo}
\email[]{yeo@mps.mpg.de}
\affiliation{Max Planck Institute for Solar System Research, Justus-von-Liebig-Weg 3, 37077 G\"ottingen, Germany.}

\author{Sami K. Solanki}
\altaffiliation{School of Space Research, Kyung Hee University, Yongjin, Gyeonggi 446-701, Korea.}
\affiliation{Max Planck Institute for Solar System Research, Justus-von-Liebig-Weg 3, 37077 G\"ottingen, Germany.}

\author{Charlotte M. Norris}
\affiliation{Blackett Laboratory, Imperial College London, South Kensington Campus, London SW7 2AZ, UK.}

\author{Benjamin Beeck}
\affiliation{Max Planck Institute for Solar System Research, Justus-von-Liebig-Weg 3, 37077 G\"ottingen, Germany.}

\author{Yvonne C. Unruh}
\affiliation{Blackett Laboratory, Imperial College London, South Kensington Campus, London SW7 2AZ, UK.}

\author{Natalie A. Krivova}
\affiliation{Max Planck Institute for Solar System Research, Justus-von-Liebig-Weg 3, 37077 G\"ottingen, Germany.}

\date{\today}

\begin{abstract}
The variation in the radiative output of the Sun, described in terms of solar irradiance, is important to climatology. A common assumption is that solar irradiance variability is driven by its surface magnetism. Verifying this assumption has, however, been hampered by the fact that models of solar irradiance variability based on solar surface magnetism have to be calibrated to observed variability. Making use of realistic three-dimensional magnetohydrodynamic simulations of the solar atmosphere and state-of-the-art solar magnetograms from the Solar Dynamics Observatory, we present a model of total solar irradiance, TSI that does not require any such calibration. In doing so, the modelled irradiance variability is entirely independent of the observational record. {(The absolute level is calibrated to the TSI record from the Total Irradiance Monitor, TIM).} The model replicates $95\%$ of the observed variability between April 2010 and July 2016, leaving little scope for alternative drivers of solar irradiance variability {at least over the timescales examined (days to years)}.
\end{abstract}

\pacs{96.60.Hv; 96.60.Ub; 96.60.Q-}

\maketitle

The brightness of the Sun is usually described in terms of solar irradiance, defined as the solar radiative flux at {1 AU}. The integral over all wavelengths, Total Solar Irradiance or TSI, has been monitored from space since 1978 \citep{kopp14}. The observation of TSI revealed fluctuations with solar activity \citep{willson81}, confirming speculations that the radiative output of the Sun is not constant but modulated by its activity \citep{eddy76}. These fluctuations, while minute (about $0.1\%$ over the 11-year solar cycle), influence the Earth's climate, with implications for our assessment of the anthropogenic contribution to climate change \citep{gray10}. Climate simulations require historical solar irradiance variability as input and this can only be provided by models since measurements only go back to 1978. Evidently, modelling solar irradiance variability requires an understanding of the underlying physics. This understanding is also vital to our interpretation of the brightness variations of other cool stars, which are beginning to be monitored extensively with missions such as COROT \citep[COnvection ROtation and planetary Transits,][]{baglin02} and Kepler \citep{gilliland10}.

The kiloGauss-strength magnetic concentrations found in the lower solar atmosphere affect the temperature structure of the enclosed plasma, forming what we observe as dark sunspots and bright faculae and network \citep{spruit83}. (Faculae refers to the bright features in magnetically active regions, which are concentrated at mid-latitudes, and network to the smaller bright features distributed more uniformly across the solar disk.) Hereafter, we will refer to faculae and network, the fundamental physics of which is similar, collectively as faculae. Observations indicate that TSI is lower and higher around the minima and maxima of the solar cycle \citep{willson88}. TSI measurements also indicate increases and decreases that coincide with the passage of faculae and sunspots across the disk \citep{willson81,hudson82,oster82}. This apparent correlation between solar irradiance variability and solar surface magnetism spurred the development of models aimed at reconstructing solar irradiance variability by ascribing it to magnetic activity on the solar surface \citep{foukal86,domingo09}. This implicitly assumes that the heat blocked by sunspots is redistributed in the convection zone and the excess heat emitted by faculae originates from the convection zone \citep{spruit82}. If the intensity deficit/excess of surface magnetic features is channeled to/from their surroundings instead, they would have no net effect on solar irradiance.

Other mechanisms have also been proposed, for example, global oscillations driven by the rotation of the Sun \citep{wolff87}, surface temperature fluctuations related to magnetic fields in the interior \citep{kuhn88}, and the solar dynamo modulating convective flows \citep{cossette13}. There has been no reported attempt to incorporate any of these mechanisms in solar irradiance models. Solar irradiance variability from acoustic oscillations, convection and flares, which occur at timescales of a day and shorter, and from the thermal and chemical evolution of the Sun, which emerge at timescales exceeding $10^5$ years, are irrelevant over the intermediate, climate-relevant timescales \citep{solanki13}.

Solar irradiance variability is modelled as the outcome of solar surface magnetism by inferring the intensity deficit and excess effected by sunspots and faculae from solar observations indicative of its surface magnetism. There are two main approaches, termed proxy and semi-empirical \citep{domingo09}. Proxy models employ indirect, disk-integrated measures of solar magnetism such as the Mg II index \citep{heath86} and the photometric sunspot index \citep{hudson82} as indications of facular brightening and sunspot darkening. Solar irradiance variability is obtained by the regression of these measures to solar irradiance observations \citep{frohlich04}. In the semi-empirical approach, solar irradiance variability is reconstructed by combining the information about the spatial distribution of solar surface magnetism in resolved full-disk observations (such as magnetograms, which map the magnetic flux density) with the intensity contrast of solar surface features calculated from models of their atmospheric structure by the solution to the radiative transfer equation \citep{fligge00}.

While such models have managed to replicate most of the variability in solar irradiance observations  \citep{solanki13}, they suffer from the shortcoming of having to rely on empirical relationships to establish facular brightening and, in the case of proxy models, sunspot darkening from the solar observations employed \citep{yeo14b}. (Semi-empirical models return sunspot darkening without taking recourse to empirical relationships.) These empirical relationships are constrained by optimizing the model output to observed solar irradiance variability. Critically, in doing so, the possible contribution to solar irradiance variability by other mechanisms can be wrongly ascribed to solar surface magnetism. The ability of present-day models to replicate observed solar irradiance variability cannot exclude the possibility that mechanisms other than solar surface magnetism might play a significant role. The true relationship between solar irradiance variability and solar surface magnetism will remain an open question for so long as models have to be calibrated to observed variability. It is therefore vital to develop models of solar irradiance that do completely without such calibrations.

Proxy models, by their very definition, have to be optimized to measured solar irradiance. Current semi-empirical models represent solar surface features with one-dimensional (1D) model atmospheres, which describe the vertical stratification of their atmospheric structure. As we will explain later in this article, the challenge in describing the variation in facular intensity with magnetic flux density \citep{yeo13} with 1D model atmospheres is the reason why existing semi-empirical models have to be calibrated to observed solar irradiance variability.

We present the first TSI model to reproduce observed variability without requiring any calibration to the latter \footnote{See Supplementary Material for a detailed description, which includes references \cite{martinezpillet97,buehler15,rogers96,anders89,vogler04,beeck13,piskunov95,bellotrubio02,rutten82,shchukina01,krivova06}}. This is achieved by extending the current semi-empirical approach to incorporate state-of-the-art 3D model atmospheres. To this end, we made use of 3D magnetohydrodynamic (MHD) simulations of the solar atmosphere generated with the MURaM code \citep{vogler05}. (MURaM denotes the \underline{M}ax-Planck Institute for Solar System Research and \underline{U}niversity of Chicago \underline{Ra}diation \underline{M}HD code.) We exploit the realism of MURaM simulations of the solar atmosphere, which have been demonstrated to reproduce a wide range of solar observations \citep{shelyag04,afram11,danilovic13,riethmuller14}. We also employed full-disc magnetograms and intensity images from the Helioseismic and Magnetic Imager onboard the Solar Dynamics Observatory, SDO/HMI, launched in 2010 \citep{schou12}. Recorded from space, HMI observations are free from atmospheric seeing effects. Also, as compared to other full-disk magnetographs, the spatial resolution is higher and the noise level is, in most cases, lower \citep{yeo14a}.

The TSI model is a major extension of the semi-empirical model SATIRE-S \citep[Spectral And Total Irradiance REconstruction for the Satellite era,][]{fligge00,yeo14a}. In SATIRE-S, the wavelength-integrated or bolometric intensity of the quiet Sun, faculae and sunspots, at various distances from disk center, is calculated on the basis of 1D model atmospheres with a radiative transfer code \citep{unruh99}. For a given day, the surface coverage by faculae and sunspots is determined by identifying these features in a full-disk magnetogram and the concurrent continuum intensity image. TSI is recovered by assigning the calculated bolometric intensities to each disk position according to this segmentation and integrating the resultant bolometric image over the solar disk. The TSI model presented here assumes a similar architecture, with the crucial exception that the bolometric intensity of the quiet Sun and faculae is calculated from 3D model atmospheres provided by MURaM simulations.

It is known that the intensity of faculae varies not just with distance from disk center, but also with magnetic flux density \citep[][and references therein]{title92,ortiz02,yeo13}. However, in SATIRE-S, facular intensity is determined as a function of distance from disk center alone with a 1D model atmosphere that corresponds to faculae of a particular and \textit{a priori} unknown magnetic flux density \citep{unruh99}. In order to relate the calculated facular intensities to measured facular magnetic flux densities, as needed to assign the appropriate intensity to observed faculae, the model introduces an empirical relationship that is constrained by optimizing the model output to measured TSI variability \citep{fligge00}. Other existing semi-empirical models face the same issue as they also employ 1D model atmospheres. The application of the calculated sunspot intensities to observed sunspots yields sunspot darkening without the need to introduce any further calibration to measured TSI variability.

To model facular brightening and therefore TSI variability without optimizing the model output to observed TSI variability, we need to incorporate the variation in facular bolometric intensity with magnetic flux density into the model in a physically self-consistent manner. This is achieved here by synthesizing bolometric images and magnetograms of 3D model atmospheres emulating how they would appear to a bolometric imager and to HMI. From the result, we determined facular bolometric intensity as a function of distance from disk center and HMI magnetogram signal. Such a quantitative relationship is exactly what is missing from existing models. This relationship allows us to assign the appropriate bolometric intensity to faculae identified in HMI magnetograms by their disc position and apparent magnetic flux density alone, making it possible to reconstruct TSI variability from HMI observations without any calibration to observed variability.

\begin{figure}
\includegraphics[width=.8\textwidth]{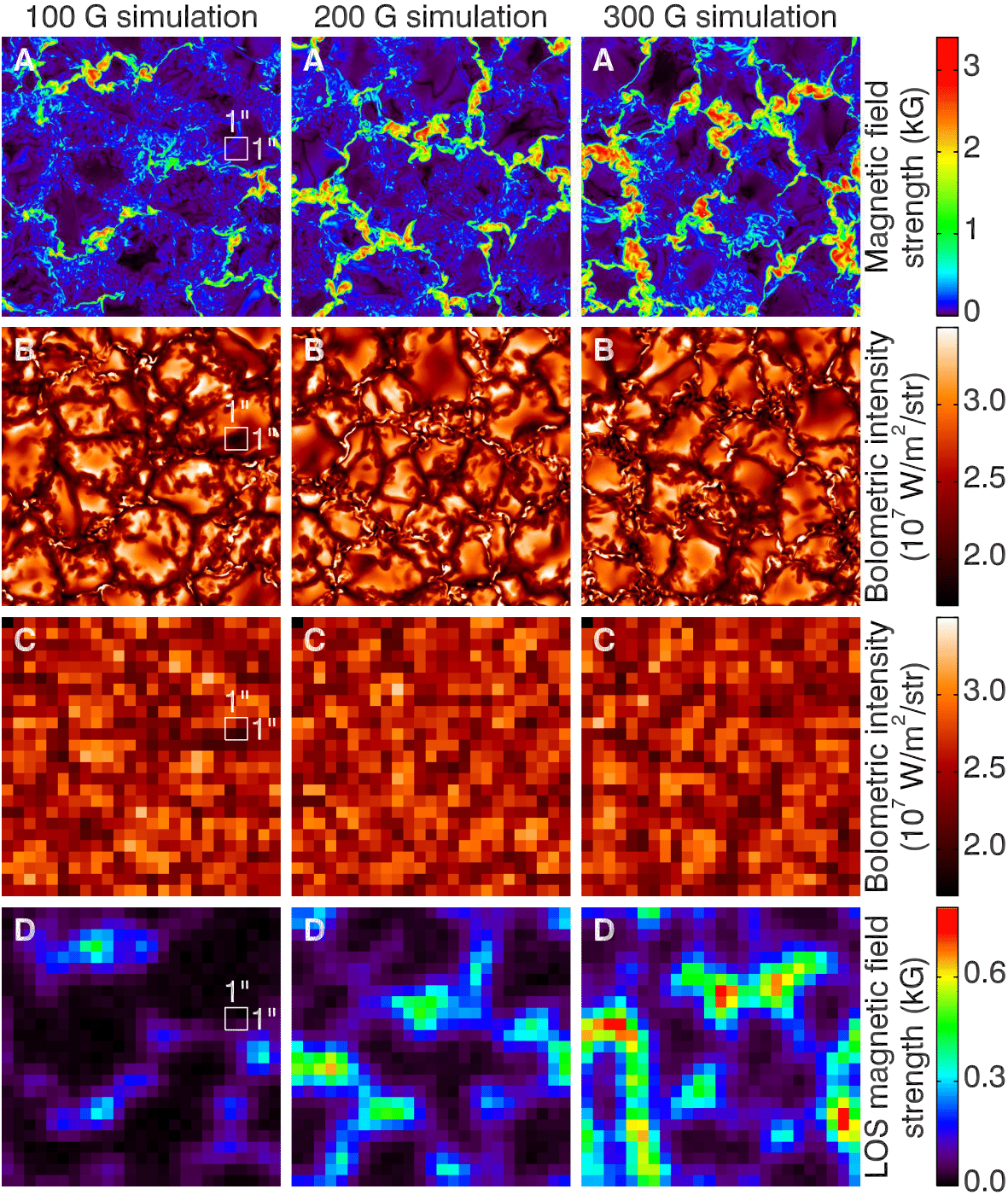}
\caption{From top to bottom: A) Magnetic field strength, at optical depth unity, in a snapshot of each MHD simulation. There are three MHD simulations, initiated with a uniform magnetic field of 100 G (left), 200 G (center) and 300 G (right), respectively. There is an additional simulation that is purely hydrodynamic (not pictured). The white box on the left panel denotes a $1"\times1"$ area. B) Bolometric image of each snapshot when viewed from above, therefore representing if they were at solar disk center, synthesized with the snapshots as input to a radiative transfer code. C) The same bolometric images, resampled to the image pixel scale of HMI. D) Corresponding longitudinal magnetogram of each snapshot, synthesized emulating the HMI instrument. The longitudinal magnetograms map the line-of-sight (LOS) component of the magnetic field strength.\label{syntheticimagesmagnetograms}}
\end{figure}

The MURaM code describes the time evolution of magnetized plasma of solar composition from magneto-convection in a 3D space \citep{vogler05}. The computational domain of the employed simulations extends $9\times9$ Mm in the horizontal and 3 Mm in the vertical, setup such that the surface of optical depth unity (i.e., the solar surface) is about 0.7 Mm from the top. We generated three MHD simulations, initiated with a uniform magnetic field of 100 G, 200 G and 300 G, respectively (Fig. \ref{syntheticimagesmagnetograms}A). There is an additional simulation to which no magnetic flux was introduced, making it purely hydrodynamic. We took 10 snapshots of each simulation. Each snapshot, capturing the state of the given simulation at a particular simulation time, is equivalent to a 3D model atmosphere.

Taking each snapshot, we calculated the emergent intensity spectrum (including the effect of spectral lines) from each horizontal position when viewed from above using a radiative transfer code \citep{kurucz92}. Taking the integral over wavelength of each intensity spectrum, we arrive at the bolometric image of the snapshot as it would appear at disk center (Fig. \ref{syntheticimagesmagnetograms}B). This is repeated, inclining the line-of-sight as needed to synthesize the bolometric images of the snapshot as it would appear at various distances from disk center. The bolometric images are resampled to the image pixel scale of HMI (Fig. \ref{syntheticimagesmagnetograms}C).

\begin{figure}
\includegraphics[width=\textwidth]{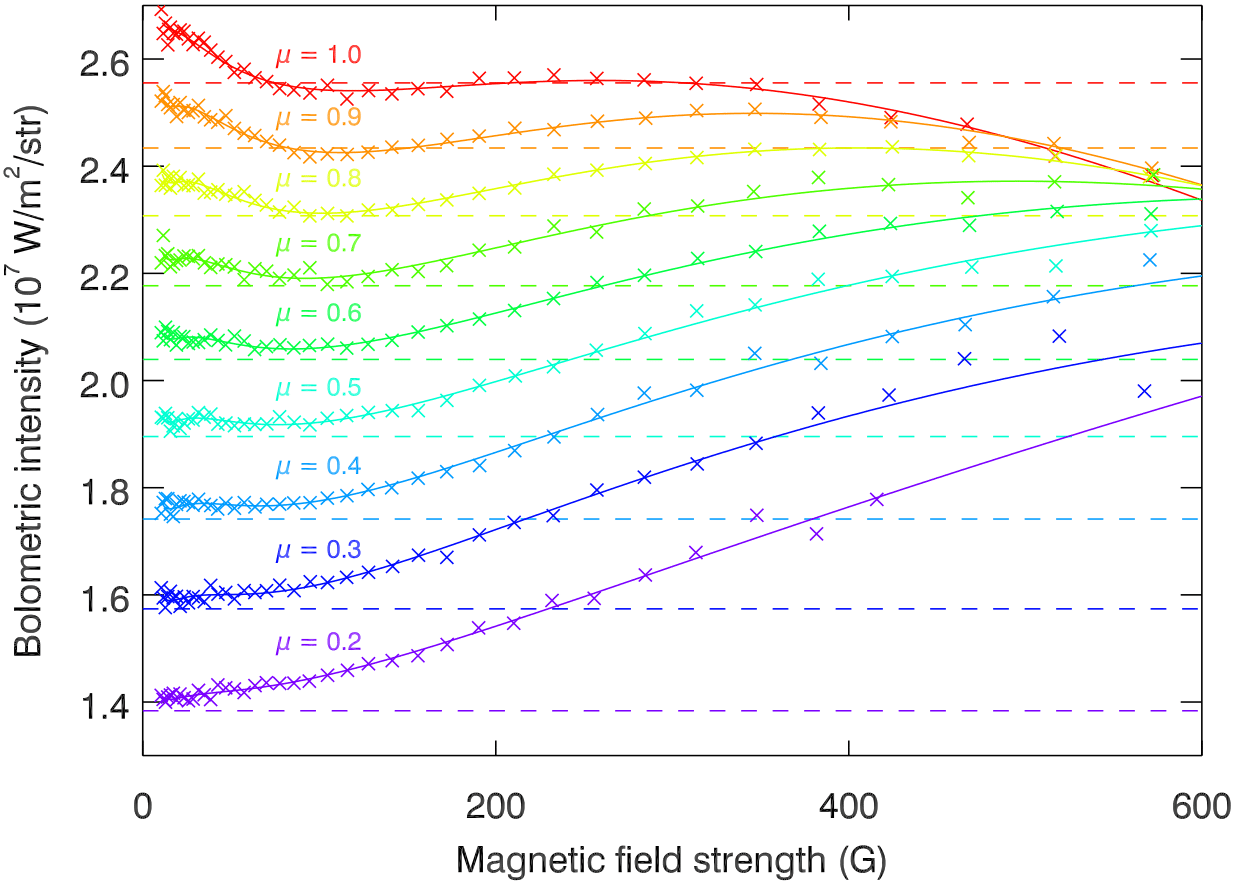}
\caption{{Bolometric intensity of the quiet Sun and faculae. Let $\mu$ denote the cosine of the heliocentric angle, representing distance from disc centre. The crosses denote the result of taking the bolometric images and the HMI-like longitudinal magnetograms of the MHD snapshots, binning the points by $\mu$ and magnetic field strength, and taking the bin-averaged magnetic field strength and bolometric intensity. The solid lines follow the corresponding bivariate polynomial in $\mu$ and magnetic field strength fit, taken into the model to describe the bolometric intensity of faculae. The dashed lines denote the quiet Sun bolometric intensity at each value of $\mu$, given by the mean level in the bolometric images of the hydrodynamic snapshots.} \label{lut}}
\end{figure}

Next, we synthesized the magnetogram corresponding to each bolometric image by emulating the HMI instrument (Fig. \ref{syntheticimagesmagnetograms}D). Applying a spectral line synthesis code \citep{frutiger00} to the MHD snapshots in a process similar to the bolometric intensity calculations just described, we calculated the Stokes parameters of the Fe I 6173 \AA{} line (the spectral line HMI observes) corresponding to each point in each bolometric image. The line profiles were processed in such a way as to emulate the spatial resolution, spectral sampling, stray light \citep{yeo14c} and noise level \citep{yeo13} of the HMI instrument. We calculated magnetic flux density from the line profiles employing the same algorithm used in the HMI data reduction pipeline \citep{couvidat12}, yielding the HMI-like magnetogram corresponding to each bolometric image. From the synthetic images and magnetograms based on the MHD snapshots, we derived a relationship describing bolometric intensity as a function of distance from disk center and HMI magnetogram signal {(solid lines, Fig. \ref{lut})}.

We determined the bolometric intensity of the quiet Sun from the bolometric images of the hydrodynamic snapshots {(dashed lines, Fig. \ref{lut})}. As in SATIRE-S, the bolometric intensity of sunspots is calculated from 1D model atmospheres. Since existing semi-empirical models already demonstrate this approach to return sunspot darkening without requiring any calibration to observed solar irradiance variability, there is no need to model it differently here.

\begin{figure}
\includegraphics[width=\textwidth]{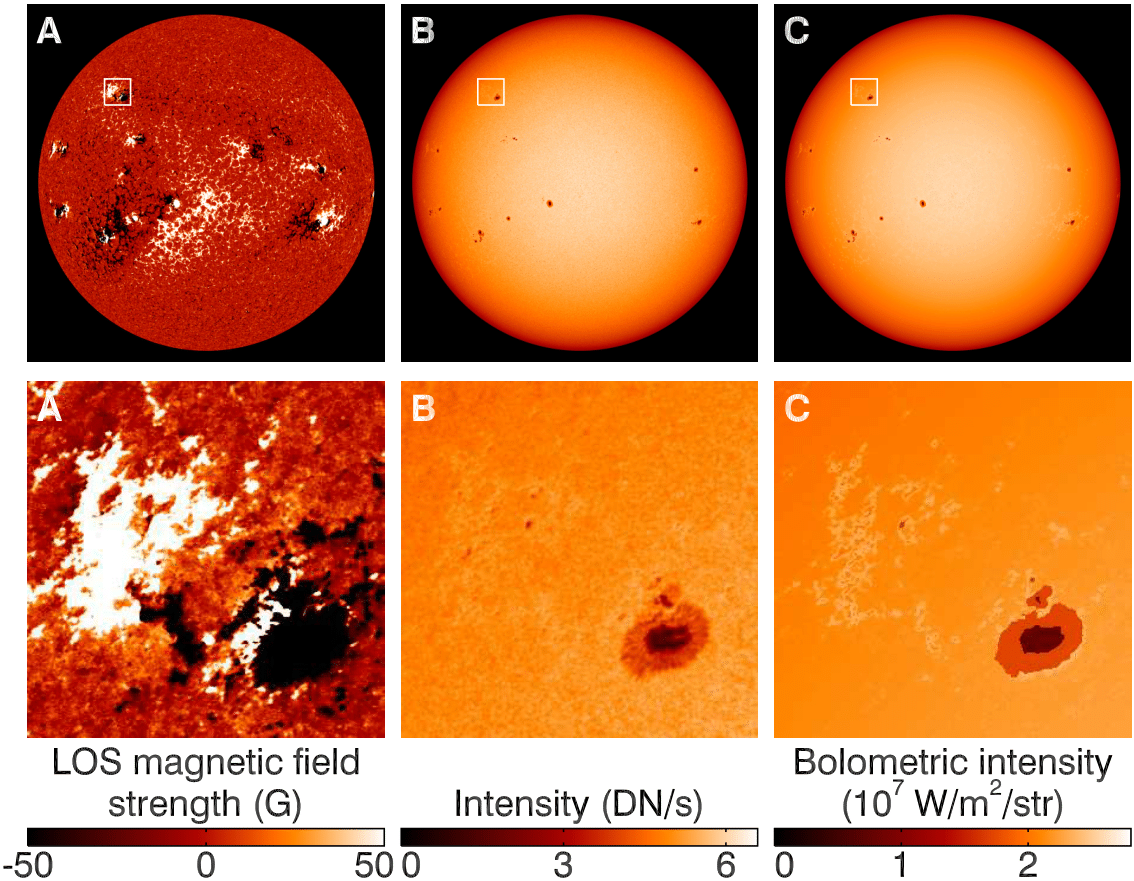}
\caption{From left to right: A) HMI longitudinal magnetogram dated December 16, 2012 (top) and the inset of the boxed area (bottom). Positive and negative values, rendered bright and dark, denote magnetic field pointed towards and away from the observer, respectively. B) Simultaneous HMI continuum intensity image, where sunspots are easily apparent. The instrument maps the intensity in the continuum of the Fe I 6173 Å line in instrumental units of digital number per second (DN/s). C) Bolometric image reconstructed with the model presented, the full-disk integral of which yields TSI. The model assigns, to each image pixel on the solar disk, the calculated bolometric intensity of either the quiet Sun, faculae, sunspot penumbra or sunspot umbra depending on what it is identified to be in the HMI observations. \label{modelbolometry}}
\end{figure}

\begin{figure}
\includegraphics[width=\textwidth]{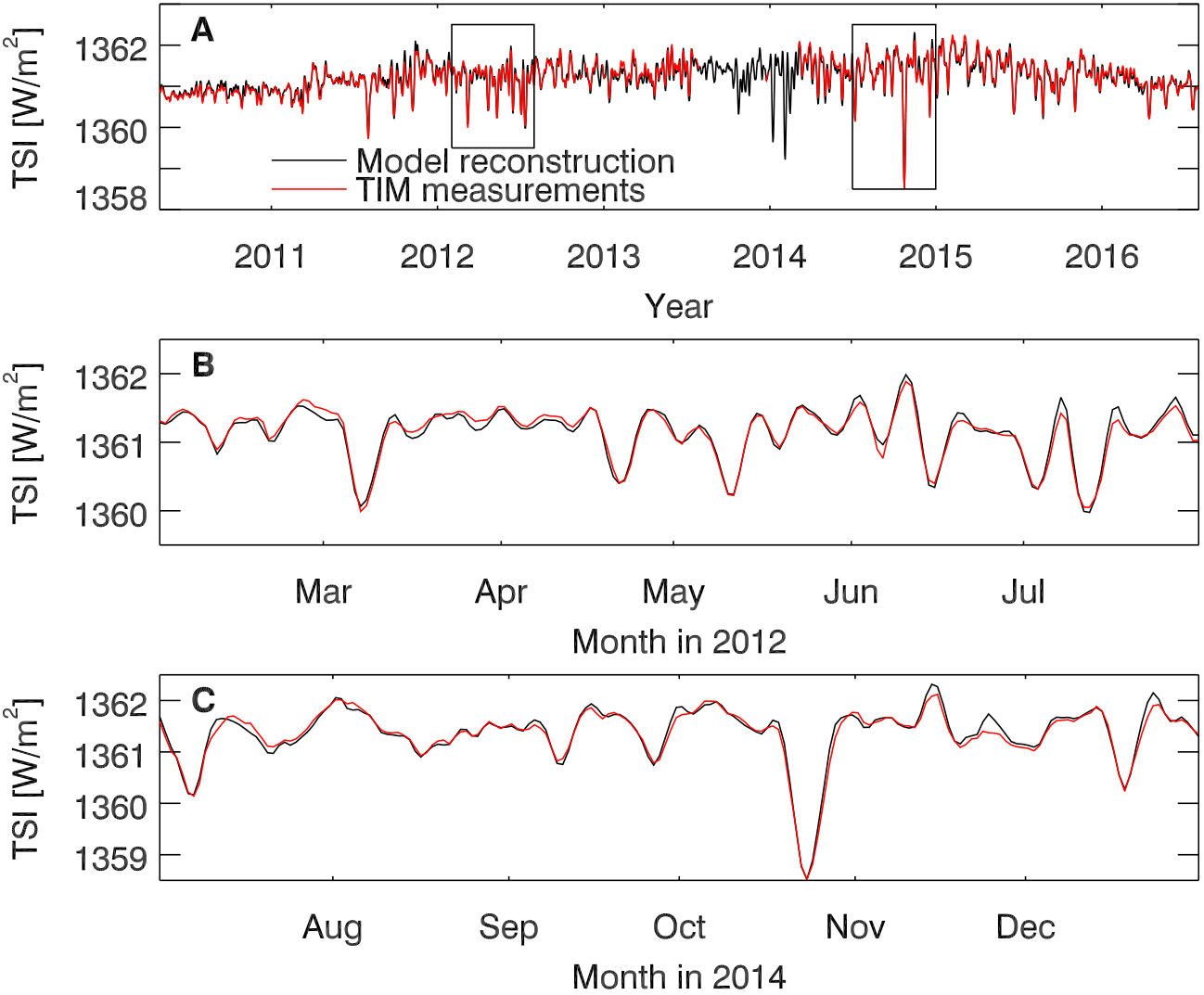}
\caption{A) The TSI reconstruction (black) and the measurements from TIM (red), which operated over the period of the reconstruction except between August 2013 and April 2014. B) The inset of the boxed period in 2012. C) The inset of the boxed period in 2014. \label{tsi}}
\end{figure}

HMI has been returning observations continuously since April 30, 2010. For each day up to July 31, 2016, we took a full-disk magnetogram (Fig. \ref{modelbolometry}A) and the concurrent continuum image (Fig. \ref{modelbolometry}B). We segmented the solar disk into the quiet Sun, faculae and sunspots by identifying faculae by the magnetogram signal and sunspots by the contrast in the continuum intensity image \citep[following][]{yeo14a}. We assigned, to each image pixel on the solar disc, the appropriate bolometric intensity by the surface feature type, distance from disk center and in the case of faculae, magnetogram signal. TSI is then given by the integral of the resultant full-disk bolometric image (Fig. \ref{modelbolometry}C). The reconstructed TSI variability (Fig. \ref{tsi}) is a direct consequence of the day-to-day variation in surface coverage by faculae and sunspots.

We compared the TSI reconstruction to the measurements from the two solar radiometers that operated over the same period, SORCE/TIM \citep{kopp05} and SoHO/VIRGO \citep{frohlich95}. (SORCE/TIM abbreviates the Total Irradiance Monitor onboard the SOlar Radiation and Climate Experiment and SOHO/VIRGO the Variability of IRradiance and Gravity Oscillations instrument onboard the SOlar and Heliospheric Observatory.) Because it is TSI variability that is of interest rather than the absolute radiative flux, we normalized the reconstruction and the VIRGO record to the TIM record, the absolute radiometry of which is widely accepted as the most accurate \citep{kopp11,kopp12,fehlmann12}.

The TSI reconstruction is in excellent agreement with both observational records, in particular that from TIM (Table 1). The Pearson's correlation coefficient, $R$ is 0.976, indicating the model replicates $95\%$ of the apparent variability. More importantly, the reconstruction reproduces the amplitude of measured TSI variability remarkably well. The reconstruction and the TIM record run closely parallel to one another, both in terms of the overall and the day-to-day trend, as visibly evident in Fig. \ref{tsi} and encapsulated in the minute root-mean-square (RMS) difference of 0.0836 ${\rm W/m^2}$.

\begin{table}
\caption{The Pearson's correlation coefficient, $R$ and RMS difference between the TSI reconstruction and the observational records from TIM and VIRGO. The reconstruction spans April 30, 2010 to July 31, 2016, a period of 2285 days, without interruption. As there are gaps in the coverage of this period by TIM and VIRGO, we compared the various time-series to one another over just the days where measurements are available from both instruments (1996 days, 87\% of the total).\label{modelvsmeasurements}}
\begin{ruledtabular}
\begin{tabular}{ccc}
Time-series	& $R$	& RMS difference (${\rm W/m^2}$) \\
\hline
Model and TIM	& 0.976	& 0.0836 \\
Model and VIRGO	& 0.968	& 0.0941 \\
TIM and VIRGO	& 0.975	& 0.0865 \\
\end{tabular}
\end{ruledtabular}
\end{table}

It is worth noting that the agreement between the model reconstruction and each of the two observational records, whether in terms of $R$ or the RMS difference, is comparable to that between the two records (Table 1). This suggests that the differences between the reconstruction and the two observational records is in no small part due to the uncertainty in the latter.

Coming without any optimization to measured TSI variability, the agreement between the TSI reconstruction and observations is direct evidence that, at least for timescales from days to the solar cycle, solar irradiance variability is mainly driven by solar surface magnetic activity. Recall, the case for this relationship has so far been circumstantial due to the fact that preceding models, without exception, have to be calibrated to observed solar irradiance variability. {The fact that the model accounts for the bulk of measured TSI variability (95\%) also implies that the contribution by other mechanisms, such as those stated earlier in this article \citep{wolff87,kuhn88,cossette13}, is very limited (no more than 5\% on the timescales examined here).}

The ability of the TSI model presented to recreate observed variability confirms the assertion that the energy blocked by sunspots is redistributed in the convection zone and the excess energy released at faculae originates from the convection zone \citep{domingo09,spruit82,solanki13}. The convection zone acts like an energy store due to the high thermal conductivity and slow thermal relaxation. Together with the high heat capacity, the fluctuations in internal energy from these exchanges have essentially no effect on the surface properties. If the intensity deficit/excess of surface magnetic features is channeled to/from their surroundings instead, accounting for just that as we have done here should not recreate observed solar irradiance variability.

Climate simulations rely on model reconstructions of historical solar irradiance variability and all current models are based on solar surface magnetism. The demonstration here that solar irradiance variability is indeed dominantly driven by solar surface magnetism bolsters the validity of their application to climate simulations.

Finally, the results of this study indicate that the photometric variability of other cool stars, at timescales greater than a day, can be almost completely ascribed to stellar surface magnetism.

\begin{acknowledgments}
We are very grateful to Yang Liu and Philip Scherrer at Stanford University for the useful discussions and their assistance in implementing the HMI observables algorithm. We made use of observations from SDO/HMI (level 1.5, available at jsoc.stanford.edu), SOHO/VIRGO (version $06\_005\_1608$, ftp.pmodwrc.ch/pub/data/) and SORCE/TIM (version 14, lasp.colorado.edu/lisird), courtesy of the respective instrument teams. The TSI reconstruction presented is available at www2.mps.mpg.de/projects/sun-climate/data.html. This work was supported by the German Federal Ministry of Education and Research (Project No. 01LG1209A), the European Research Council (ERC) under the European Union’s Horizon 2020 research and innovation program (Grant Agreement No. 695075) and the BK21 plus program of the National Research Foundation (NRF) funded by the Ministry of Education of Korea.
\end{acknowledgments}


\begin{thebibliography}{53}%
\makeatletter
\providecommand \@ifxundefined [1]{%
 \@ifx{#1\undefined}
}%
\providecommand \@ifnum [1]{%
 \ifnum #1\expandafter \@firstoftwo
 \else \expandafter \@secondoftwo
 \fi
}%
\providecommand \@ifx [1]{%
 \ifx #1\expandafter \@firstoftwo
 \else \expandafter \@secondoftwo
 \fi
}%
\providecommand \natexlab [1]{#1}%
\providecommand \enquote  [1]{``#1''}%
\providecommand \bibnamefont  [1]{#1}%
\providecommand \bibfnamefont [1]{#1}%
\providecommand \citenamefont [1]{#1}%
\providecommand \href@noop [0]{\@secondoftwo}%
\providecommand \href [0]{\begingroup \@sanitize@url \@href}%
\providecommand \@href[1]{\@@startlink{#1}\@@href}%
\providecommand \@@href[1]{\endgroup#1\@@endlink}%
\providecommand \@sanitize@url [0]{\catcode `\\12\catcode `\$12\catcode
  `\&12\catcode `\#12\catcode `\^12\catcode `\_12\catcode `\%12\relax}%
\providecommand \@@startlink[1]{}%
\providecommand \@@endlink[0]{}%
\providecommand \url  [0]{\begingroup\@sanitize@url \@url }%
\providecommand \@url [1]{\endgroup\@href {#1}{\urlprefix }}%
\providecommand \urlprefix  [0]{URL }%
\providecommand \Eprint [0]{\href }%
\providecommand \doibase [0]{http://dx.doi.org/}%
\providecommand \selectlanguage [0]{\@gobble}%
\providecommand \bibinfo  [0]{\@secondoftwo}%
\providecommand \bibfield  [0]{\@secondoftwo}%
\providecommand \translation [1]{[#1]}%
\providecommand \BibitemOpen [0]{}%
\providecommand \bibitemStop [0]{}%
\providecommand \bibitemNoStop [0]{.\EOS\space}%
\providecommand \EOS [0]{\spacefactor3000\relax}%
\providecommand \BibitemShut  [1]{\csname bibitem#1\endcsname}%
\let\auto@bib@innerbib\@empty
\bibitem [{\citenamefont {{Kopp}}(2014)}]{kopp14}%
  \BibitemOpen
  \bibfield  {author} {\bibinfo {author} {\bibfnamefont {G.}~\bibnamefont
  {{Kopp}}},\ }\href {\doibase 10.1051/swsc/2014012} {\bibfield  {journal}
  {\bibinfo  {journal} {J. Space Weather Space Clim.}\ }\textbf {\bibinfo
  {volume} {4}},\ \bibinfo {pages} {A14} (\bibinfo {year} {2014})}\BibitemShut
  {NoStop}%
\bibitem [{\citenamefont {{Willson}}\ \emph {et~al.}(1981)\citenamefont
  {{Willson}}, \citenamefont {{Gulkis}}, \citenamefont {{Janssen}},
  \citenamefont {{Hudson}},\ and\ \citenamefont {{Chapman}}}]{willson81}%
  \BibitemOpen
  \bibfield  {author} {\bibinfo {author} {\bibfnamefont {R.~C.}\ \bibnamefont
  {{Willson}}}, \bibinfo {author} {\bibfnamefont {S.}~\bibnamefont {{Gulkis}}},
  \bibinfo {author} {\bibfnamefont {M.}~\bibnamefont {{Janssen}}}, \bibinfo
  {author} {\bibfnamefont {H.~S.}\ \bibnamefont {{Hudson}}}, \ and\ \bibinfo
  {author} {\bibfnamefont {G.~A.}\ \bibnamefont {{Chapman}}},\ }\href {\doibase
  10.1126/science.211.4483.700} {\bibfield  {journal} {\bibinfo  {journal}
  {Science}\ }\textbf {\bibinfo {volume} {211}},\ \bibinfo {pages} {700}
  (\bibinfo {year} {1981})}\BibitemShut {NoStop}%
\bibitem [{\citenamefont {{Eddy}}(1976)}]{eddy76}%
  \BibitemOpen
  \bibfield  {author} {\bibinfo {author} {\bibfnamefont {J.~A.}\ \bibnamefont
  {{Eddy}}},\ }\href {\doibase 10.1126/science.192.4245.1189} {\bibfield
  {journal} {\bibinfo  {journal} {Science}\ }\textbf {\bibinfo {volume}
  {192}},\ \bibinfo {pages} {1189} (\bibinfo {year} {1976})}\BibitemShut
  {NoStop}%
\bibitem [{\citenamefont {{Gray}}\ \emph {et~al.}(2010)\citenamefont {{Gray}},
  \citenamefont {{Beer}}, \citenamefont {{Geller}}, \citenamefont {{Haigh}},
  \citenamefont {{Lockwood}}, \citenamefont {{Matthes}}, \citenamefont
  {{Cubasch}}, \citenamefont {{Fleitmann}}, \citenamefont {{Harrison}},
  \citenamefont {{Hood}}, \citenamefont {{Luterbacher}}, \citenamefont
  {{Meehl}}, \citenamefont {{Shindell}}, \citenamefont {{van Geel}},\ and\
  \citenamefont {{White}}}]{gray10}%
  \BibitemOpen
  \bibfield  {author} {\bibinfo {author} {\bibfnamefont {L.~J.}\ \bibnamefont
  {{Gray}}}, \bibinfo {author} {\bibfnamefont {J.}~\bibnamefont {{Beer}}},
  \bibinfo {author} {\bibfnamefont {M.}~\bibnamefont {{Geller}}}, \bibinfo
  {author} {\bibfnamefont {J.~D.}\ \bibnamefont {{Haigh}}}, \bibinfo {author}
  {\bibfnamefont {M.}~\bibnamefont {{Lockwood}}}, \bibinfo {author}
  {\bibfnamefont {K.}~\bibnamefont {{Matthes}}}, \bibinfo {author}
  {\bibfnamefont {U.}~\bibnamefont {{Cubasch}}}, \bibinfo {author}
  {\bibfnamefont {D.}~\bibnamefont {{Fleitmann}}}, \bibinfo {author}
  {\bibfnamefont {G.}~\bibnamefont {{Harrison}}}, \bibinfo {author}
  {\bibfnamefont {L.}~\bibnamefont {{Hood}}}, \bibinfo {author} {\bibfnamefont
  {J.}~\bibnamefont {{Luterbacher}}}, \bibinfo {author} {\bibfnamefont {G.~A.}\
  \bibnamefont {{Meehl}}}, \bibinfo {author} {\bibfnamefont {D.}~\bibnamefont
  {{Shindell}}}, \bibinfo {author} {\bibfnamefont {B.}~\bibnamefont {{van
  Geel}}}, \ and\ \bibinfo {author} {\bibfnamefont {W.}~\bibnamefont
  {{White}}},\ }\href {\doibase 10.1029/2009RG000282} {\bibfield  {journal}
  {\bibinfo  {journal} {Rev. Geophys.}\ }\textbf {\bibinfo {volume} {48}},\
  \bibinfo {pages} {RG4001} (\bibinfo {year} {2010})}\BibitemShut {NoStop}%
\bibitem [{\citenamefont {{Baglin}}\ \emph {et~al.}(2002)\citenamefont
  {{Baglin}}, \citenamefont {{Auvergne}}, \citenamefont {{Barge}},
  \citenamefont {{Buey}}, \citenamefont {{Catala}}, \citenamefont {{Michel}},
  \citenamefont {{Weiss}},\ and\ \citenamefont {{COROT Team}}}]{baglin02}%
  \BibitemOpen
  \bibfield  {author} {\bibinfo {author} {\bibfnamefont {A.}~\bibnamefont
  {{Baglin}}}, \bibinfo {author} {\bibfnamefont {M.}~\bibnamefont
  {{Auvergne}}}, \bibinfo {author} {\bibfnamefont {P.}~\bibnamefont {{Barge}}},
  \bibinfo {author} {\bibfnamefont {J.-T.}\ \bibnamefont {{Buey}}}, \bibinfo
  {author} {\bibfnamefont {C.}~\bibnamefont {{Catala}}}, \bibinfo {author}
  {\bibfnamefont {E.}~\bibnamefont {{Michel}}}, \bibinfo {author}
  {\bibfnamefont {W.}~\bibnamefont {{Weiss}}}, \ and\ \bibinfo {author}
  {\bibnamefont {{COROT Team}}},\ }in\ \href@noop {} {\emph {\bibinfo
  {booktitle} {Stellar Structure and Habitable Planet Finding}}},\ \bibinfo
  {series} {ESA Spec. Publ.}, Vol.\ \bibinfo {volume} {485},\ \bibinfo {editor}
  {edited by\ \bibinfo {editor} {\bibfnamefont {B.}~\bibnamefont {{Battrick}}},
  \bibinfo {editor} {\bibfnamefont {F.}~\bibnamefont {{Favata}}}, \bibinfo
  {editor} {\bibfnamefont {I.~W.}\ \bibnamefont {{Roxburgh}}}, \ and\ \bibinfo
  {editor} {\bibfnamefont {D.}~\bibnamefont {{Galadi}}}}\ (\bibinfo {year}
  {2002})\ pp.\ \bibinfo {pages} {17--24}\BibitemShut {NoStop}%
\bibitem [{\citenamefont {{Gilliland}}\ \emph {et~al.}(2010)\citenamefont
  {{Gilliland}}, \citenamefont {{Brown}}, \citenamefont
  {{Christensen-Dalsgaard}}, \citenamefont {{Kjeldsen}}, \citenamefont
  {{Aerts}}, \citenamefont {{Appourchaux}}, \citenamefont {{Basu}},
  \citenamefont {{Bedding}}, \citenamefont {{Chaplin}}, \citenamefont
  {{Cunha}}, \citenamefont {{De Cat}}, \citenamefont {{De Ridder}},
  \citenamefont {{Guzik}}, \citenamefont {{Handler}}, \citenamefont
  {{Kawaler}}, \citenamefont {{Kiss}}, \citenamefont {{Kolenberg}},
  \citenamefont {{Kurtz}}, \citenamefont {{Metcalfe}}, \citenamefont
  {{Monteiro}}, \citenamefont {{Szab{\'o}}}, \citenamefont {{Arentoft}},
  \citenamefont {{Balona}}, \citenamefont {{Debosscher}}, \citenamefont
  {{Elsworth}}, \citenamefont {{Quirion}}, \citenamefont {{Stello}},
  \citenamefont {{Su{\'a}rez}}, \citenamefont {{Borucki}}, \citenamefont
  {{Jenkins}}, \citenamefont {{Koch}}, \citenamefont {{Kondo}}, \citenamefont
  {{Latham}}, \citenamefont {{Rowe}},\ and\ \citenamefont
  {{Steffen}}}]{gilliland10}%
  \BibitemOpen
  \bibfield  {author} {\bibinfo {author} {\bibfnamefont {R.~L.}\ \bibnamefont
  {{Gilliland}}}, \bibinfo {author} {\bibfnamefont {T.~M.}\ \bibnamefont
  {{Brown}}}, \bibinfo {author} {\bibfnamefont {J.}~\bibnamefont
  {{Christensen-Dalsgaard}}}, \bibinfo {author} {\bibfnamefont
  {H.}~\bibnamefont {{Kjeldsen}}}, \bibinfo {author} {\bibfnamefont
  {C.}~\bibnamefont {{Aerts}}}, \bibinfo {author} {\bibfnamefont
  {T.}~\bibnamefont {{Appourchaux}}}, \bibinfo {author} {\bibfnamefont
  {S.}~\bibnamefont {{Basu}}}, \bibinfo {author} {\bibfnamefont {T.~R.}\
  \bibnamefont {{Bedding}}}, \bibinfo {author} {\bibfnamefont {W.~J.}\
  \bibnamefont {{Chaplin}}}, \bibinfo {author} {\bibfnamefont {M.~S.}\
  \bibnamefont {{Cunha}}}, \bibinfo {author} {\bibfnamefont {P.}~\bibnamefont
  {{De Cat}}}, \bibinfo {author} {\bibfnamefont {J.}~\bibnamefont {{De
  Ridder}}}, \bibinfo {author} {\bibfnamefont {J.~A.}\ \bibnamefont {{Guzik}}},
  \bibinfo {author} {\bibfnamefont {G.}~\bibnamefont {{Handler}}}, \bibinfo
  {author} {\bibfnamefont {S.}~\bibnamefont {{Kawaler}}}, \bibinfo {author}
  {\bibfnamefont {L.}~\bibnamefont {{Kiss}}}, \bibinfo {author} {\bibfnamefont
  {K.}~\bibnamefont {{Kolenberg}}}, \bibinfo {author} {\bibfnamefont {D.~W.}\
  \bibnamefont {{Kurtz}}}, \bibinfo {author} {\bibfnamefont {T.~S.}\
  \bibnamefont {{Metcalfe}}}, \bibinfo {author} {\bibfnamefont
  {M.~J.~P.~F.~G.}\ \bibnamefont {{Monteiro}}}, \bibinfo {author}
  {\bibfnamefont {R.}~\bibnamefont {{Szab{\'o}}}}, \bibinfo {author}
  {\bibfnamefont {T.}~\bibnamefont {{Arentoft}}}, \bibinfo {author}
  {\bibfnamefont {L.}~\bibnamefont {{Balona}}}, \bibinfo {author}
  {\bibfnamefont {J.}~\bibnamefont {{Debosscher}}}, \bibinfo {author}
  {\bibfnamefont {Y.~P.}\ \bibnamefont {{Elsworth}}}, \bibinfo {author}
  {\bibfnamefont {P.-O.}\ \bibnamefont {{Quirion}}}, \bibinfo {author}
  {\bibfnamefont {D.}~\bibnamefont {{Stello}}}, \bibinfo {author}
  {\bibfnamefont {J.~C.}\ \bibnamefont {{Su{\'a}rez}}}, \bibinfo {author}
  {\bibfnamefont {W.~J.}\ \bibnamefont {{Borucki}}}, \bibinfo {author}
  {\bibfnamefont {J.~M.}\ \bibnamefont {{Jenkins}}}, \bibinfo {author}
  {\bibfnamefont {D.}~\bibnamefont {{Koch}}}, \bibinfo {author} {\bibfnamefont
  {Y.}~\bibnamefont {{Kondo}}}, \bibinfo {author} {\bibfnamefont {D.~W.}\
  \bibnamefont {{Latham}}}, \bibinfo {author} {\bibfnamefont {J.~F.}\
  \bibnamefont {{Rowe}}}, \ and\ \bibinfo {author} {\bibfnamefont {J.~H.}\
  \bibnamefont {{Steffen}}},\ }\href {\doibase 10.1086/650399} {\bibfield
  {journal} {\bibinfo  {journal} {Pub. Astron. Soc. Pac.}\ }\textbf {\bibinfo
  {volume} {122}},\ \bibinfo {pages} {131} (\bibinfo {year}
  {2010})}\BibitemShut {NoStop}%
\bibitem [{\citenamefont {{Spruit}}\ and\ \citenamefont
  {{Roberts}}(1983)}]{spruit83}%
  \BibitemOpen
  \bibfield  {author} {\bibinfo {author} {\bibfnamefont {H.~C.}\ \bibnamefont
  {{Spruit}}}\ and\ \bibinfo {author} {\bibfnamefont {B.}~\bibnamefont
  {{Roberts}}},\ }\href {\doibase 10.1038/304401a0} {\bibfield  {journal}
  {\bibinfo  {journal} {Nature}\ }\textbf {\bibinfo {volume} {304}},\ \bibinfo
  {pages} {401} (\bibinfo {year} {1983})}\BibitemShut {NoStop}%
\bibitem [{\citenamefont {{Willson}}\ and\ \citenamefont
  {{Hudson}}(1988)}]{willson88}%
  \BibitemOpen
  \bibfield  {author} {\bibinfo {author} {\bibfnamefont {R.~C.}\ \bibnamefont
  {{Willson}}}\ and\ \bibinfo {author} {\bibfnamefont {H.~S.}\ \bibnamefont
  {{Hudson}}},\ }\href {\doibase 10.1038/332810a0} {\bibfield  {journal}
  {\bibinfo  {journal} {Nature}\ }\textbf {\bibinfo {volume} {332}},\ \bibinfo
  {pages} {810} (\bibinfo {year} {1988})}\BibitemShut {NoStop}%
\bibitem [{\citenamefont {{Hudson}}\ \emph {et~al.}(1982)\citenamefont
  {{Hudson}}, \citenamefont {{Silva}}, \citenamefont {{Woodard}},\ and\
  \citenamefont {{Willson}}}]{hudson82}%
  \BibitemOpen
  \bibfield  {author} {\bibinfo {author} {\bibfnamefont {H.~S.}\ \bibnamefont
  {{Hudson}}}, \bibinfo {author} {\bibfnamefont {S.}~\bibnamefont {{Silva}}},
  \bibinfo {author} {\bibfnamefont {M.}~\bibnamefont {{Woodard}}}, \ and\
  \bibinfo {author} {\bibfnamefont {R.~C.}\ \bibnamefont {{Willson}}},\ }\href
  {\doibase 10.1007/BF00170984} {\bibfield  {journal} {\bibinfo  {journal}
  {Sol. Phys.}\ }\textbf {\bibinfo {volume} {76}},\ \bibinfo {pages} {211}
  (\bibinfo {year} {1982})}\BibitemShut {NoStop}%
\bibitem [{\citenamefont {{Oster}}\ \emph {et~al.}(1982)\citenamefont
  {{Oster}}, \citenamefont {{Schatten}},\ and\ \citenamefont
  {{Sofia}}}]{oster82}%
  \BibitemOpen
  \bibfield  {author} {\bibinfo {author} {\bibfnamefont {L.}~\bibnamefont
  {{Oster}}}, \bibinfo {author} {\bibfnamefont {K.~H.}\ \bibnamefont
  {{Schatten}}}, \ and\ \bibinfo {author} {\bibfnamefont {S.}~\bibnamefont
  {{Sofia}}},\ }\href {\doibase 10.1086/159949} {\bibfield  {journal} {\bibinfo
   {journal} {Astrophys. J.}\ }\textbf {\bibinfo {volume} {256}},\ \bibinfo
  {pages} {768} (\bibinfo {year} {1982})}\BibitemShut {NoStop}%
\bibitem [{\citenamefont {{Foukal}}\ and\ \citenamefont
  {{Lean}}(1986)}]{foukal86}%
  \BibitemOpen
  \bibfield  {author} {\bibinfo {author} {\bibfnamefont {P.}~\bibnamefont
  {{Foukal}}}\ and\ \bibinfo {author} {\bibfnamefont {J.}~\bibnamefont
  {{Lean}}},\ }\href {\doibase 10.1086/164043} {\bibfield  {journal} {\bibinfo
  {journal} {Astrophys. J. Lett.}\ }\textbf {\bibinfo {volume} {302}},\
  \bibinfo {pages} {826} (\bibinfo {year} {1986})}\BibitemShut {NoStop}%
\bibitem [{\citenamefont {{Domingo}}\ \emph {et~al.}(2009)\citenamefont
  {{Domingo}}, \citenamefont {{Ermolli}}, \citenamefont {{Fox}}, \citenamefont
  {{Fr{\"o}hlich}}, \citenamefont {{Haberreiter}}, \citenamefont {{Krivova}},
  \citenamefont {{Kopp}}, \citenamefont {{Schmutz}}, \citenamefont {{Solanki}},
  \citenamefont {{Spruit}}, \citenamefont {{Unruh}},\ and\ \citenamefont
  {{V{\"o}gler}}}]{domingo09}%
  \BibitemOpen
  \bibfield  {author} {\bibinfo {author} {\bibfnamefont {V.}~\bibnamefont
  {{Domingo}}}, \bibinfo {author} {\bibfnamefont {I.}~\bibnamefont
  {{Ermolli}}}, \bibinfo {author} {\bibfnamefont {P.}~\bibnamefont {{Fox}}},
  \bibinfo {author} {\bibfnamefont {C.}~\bibnamefont {{Fr{\"o}hlich}}},
  \bibinfo {author} {\bibfnamefont {M.}~\bibnamefont {{Haberreiter}}}, \bibinfo
  {author} {\bibfnamefont {N.}~\bibnamefont {{Krivova}}}, \bibinfo {author}
  {\bibfnamefont {G.}~\bibnamefont {{Kopp}}}, \bibinfo {author} {\bibfnamefont
  {W.}~\bibnamefont {{Schmutz}}}, \bibinfo {author} {\bibfnamefont {S.~K.}\
  \bibnamefont {{Solanki}}}, \bibinfo {author} {\bibfnamefont {H.~C.}\
  \bibnamefont {{Spruit}}}, \bibinfo {author} {\bibfnamefont {Y.}~\bibnamefont
  {{Unruh}}}, \ and\ \bibinfo {author} {\bibfnamefont {A.}~\bibnamefont
  {{V{\"o}gler}}},\ }\href {\doibase 10.1007/s11214-009-9562-1} {\bibfield
  {journal} {\bibinfo  {journal} {Space Sci. Rev.}\ }\textbf {\bibinfo {volume}
  {145}},\ \bibinfo {pages} {337} (\bibinfo {year} {2009})}\BibitemShut
  {NoStop}%
\bibitem [{\citenamefont {{Spruit}}(1982)}]{spruit82}%
  \BibitemOpen
  \bibfield  {author} {\bibinfo {author} {\bibfnamefont {H.~C.}\ \bibnamefont
  {{Spruit}}},\ }\href@noop {} {\bibfield  {journal} {\bibinfo  {journal}
  {Astron. Astrophys.}\ }\textbf {\bibinfo {volume} {108}},\ \bibinfo {pages}
  {356} (\bibinfo {year} {1982})}\BibitemShut {NoStop}%
\bibitem [{\citenamefont {{Wolff}}\ and\ \citenamefont
  {{Hickey}}(1987)}]{wolff87}%
  \BibitemOpen
  \bibfield  {author} {\bibinfo {author} {\bibfnamefont {C.~L.}\ \bibnamefont
  {{Wolff}}}\ and\ \bibinfo {author} {\bibfnamefont {J.~R.}\ \bibnamefont
  {{Hickey}}},\ }\href {\doibase 10.1126/science.235.4796.1631} {\bibfield
  {journal} {\bibinfo  {journal} {Science}\ }\textbf {\bibinfo {volume}
  {235}},\ \bibinfo {pages} {1631} (\bibinfo {year} {1987})}\BibitemShut
  {NoStop}%
\bibitem [{\citenamefont {{Kuhn}}\ \emph {et~al.}(1988)\citenamefont {{Kuhn}},
  \citenamefont {{Libbrecht}},\ and\ \citenamefont {{Dicke}}}]{kuhn88}%
  \BibitemOpen
  \bibfield  {author} {\bibinfo {author} {\bibfnamefont {J.~R.}\ \bibnamefont
  {{Kuhn}}}, \bibinfo {author} {\bibfnamefont {K.~G.}\ \bibnamefont
  {{Libbrecht}}}, \ and\ \bibinfo {author} {\bibfnamefont {R.~H.}\ \bibnamefont
  {{Dicke}}},\ }\href {\doibase 10.1126/science.242.4880.908} {\bibfield
  {journal} {\bibinfo  {journal} {Science}\ }\textbf {\bibinfo {volume}
  {242}},\ \bibinfo {pages} {908} (\bibinfo {year} {1988})}\BibitemShut
  {NoStop}%
\bibitem [{\citenamefont {{Cossette}}\ \emph {et~al.}(2013)\citenamefont
  {{Cossette}}, \citenamefont {{Charbonneau}},\ and\ \citenamefont
  {{Smolarkiewicz}}}]{cossette13}%
  \BibitemOpen
  \bibfield  {author} {\bibinfo {author} {\bibfnamefont {J.-F.}\ \bibnamefont
  {{Cossette}}}, \bibinfo {author} {\bibfnamefont {P.}~\bibnamefont
  {{Charbonneau}}}, \ and\ \bibinfo {author} {\bibfnamefont {P.~K.}\
  \bibnamefont {{Smolarkiewicz}}},\ }\href {\doibase
  10.1088/2041-8205/777/2/L29} {\bibfield  {journal} {\bibinfo  {journal}
  {Astrophys. J. Lett.}\ }\textbf {\bibinfo {volume} {777}},\ \bibinfo {pages}
  {L29} (\bibinfo {year} {2013})}\BibitemShut {NoStop}%
\bibitem [{\citenamefont {{Solanki}}\ \emph {et~al.}(2013)\citenamefont
  {{Solanki}}, \citenamefont {{Krivova}},\ and\ \citenamefont
  {{Haigh}}}]{solanki13}%
  \BibitemOpen
  \bibfield  {author} {\bibinfo {author} {\bibfnamefont {S.~K.}\ \bibnamefont
  {{Solanki}}}, \bibinfo {author} {\bibfnamefont {N.~A.}\ \bibnamefont
  {{Krivova}}}, \ and\ \bibinfo {author} {\bibfnamefont {J.~D.}\ \bibnamefont
  {{Haigh}}},\ }\href {\doibase 10.1146/annurev-astro-082812-141007} {\bibfield
   {journal} {\bibinfo  {journal} {Annu. Rev. Astron. Astrophys.}\ }\textbf
  {\bibinfo {volume} {51}},\ \bibinfo {pages} {311} (\bibinfo {year}
  {2013})}\BibitemShut {NoStop}%
\bibitem [{\citenamefont {{Heath}}\ and\ \citenamefont
  {{Schlesinger}}(1986)}]{heath86}%
  \BibitemOpen
  \bibfield  {author} {\bibinfo {author} {\bibfnamefont {D.~F.}\ \bibnamefont
  {{Heath}}}\ and\ \bibinfo {author} {\bibfnamefont {B.~M.}\ \bibnamefont
  {{Schlesinger}}},\ }\href {\doibase 10.1029/JD091iD08p08672} {\bibfield
  {journal} {\bibinfo  {journal} {J. Geophys. Res.}\ }\textbf {\bibinfo
  {volume} {91}},\ \bibinfo {pages} {8672} (\bibinfo {year}
  {1986})}\BibitemShut {NoStop}%
\bibitem [{\citenamefont {{Fr{\"o}hlich}}\ and\ \citenamefont
  {{Lean}}(2004)}]{frohlich04}%
  \BibitemOpen
  \bibfield  {author} {\bibinfo {author} {\bibfnamefont {C.}~\bibnamefont
  {{Fr{\"o}hlich}}}\ and\ \bibinfo {author} {\bibfnamefont {J.}~\bibnamefont
  {{Lean}}},\ }\href {\doibase 10.1007/s00159-004-0024-1} {\bibfield  {journal}
  {\bibinfo  {journal} {Astron. Astrophys. Rev.}\ }\textbf {\bibinfo {volume}
  {12}},\ \bibinfo {pages} {273} (\bibinfo {year} {2004})}\BibitemShut
  {NoStop}%
\bibitem [{\citenamefont {{Fligge}}\ \emph {et~al.}(2000)\citenamefont
  {{Fligge}}, \citenamefont {{Solanki}},\ and\ \citenamefont
  {{Unruh}}}]{fligge00}%
  \BibitemOpen
  \bibfield  {author} {\bibinfo {author} {\bibfnamefont {M.}~\bibnamefont
  {{Fligge}}}, \bibinfo {author} {\bibfnamefont {S.~K.}\ \bibnamefont
  {{Solanki}}}, \ and\ \bibinfo {author} {\bibfnamefont {Y.~C.}\ \bibnamefont
  {{Unruh}}},\ }\href@noop {} {\bibfield  {journal} {\bibinfo  {journal}
  {Astron. Astrophys.}\ }\textbf {\bibinfo {volume} {353}},\ \bibinfo {pages}
  {380} (\bibinfo {year} {2000})}\BibitemShut {NoStop}%
\bibitem [{\citenamefont {{Yeo}}\ \emph
  {et~al.}(2014{\natexlab{a}})\citenamefont {{Yeo}}, \citenamefont
  {{Krivova}},\ and\ \citenamefont {{Solanki}}}]{yeo14b}%
  \BibitemOpen
  \bibfield  {author} {\bibinfo {author} {\bibfnamefont {K.~L.}\ \bibnamefont
  {{Yeo}}}, \bibinfo {author} {\bibfnamefont {N.~A.}\ \bibnamefont
  {{Krivova}}}, \ and\ \bibinfo {author} {\bibfnamefont {S.~K.}\ \bibnamefont
  {{Solanki}}},\ }\href {\doibase 10.1007/s11214-014-0061-7} {\bibfield
  {journal} {\bibinfo  {journal} {Space Sci. Rev.}\ }\textbf {\bibinfo {volume}
  {186}},\ \bibinfo {pages} {137} (\bibinfo {year}
  {2014}{\natexlab{a}})}\BibitemShut {NoStop}%
\bibitem [{\citenamefont {{Yeo}}\ \emph {et~al.}(2013)\citenamefont {{Yeo}},
  \citenamefont {{Solanki}},\ and\ \citenamefont {{Krivova}}}]{yeo13}%
  \BibitemOpen
  \bibfield  {author} {\bibinfo {author} {\bibfnamefont {K.~L.}\ \bibnamefont
  {{Yeo}}}, \bibinfo {author} {\bibfnamefont {S.~K.}\ \bibnamefont
  {{Solanki}}}, \ and\ \bibinfo {author} {\bibfnamefont {N.~A.}\ \bibnamefont
  {{Krivova}}},\ }\href {\doibase 10.1051/0004-6361/201220682} {\bibfield
  {journal} {\bibinfo  {journal} {Astron. Astrophys.}\ }\textbf {\bibinfo
  {volume} {550}},\ \bibinfo {pages} {A95} (\bibinfo {year}
  {2013})}\BibitemShut {NoStop}%
\bibitem [{Note1()}]{Note1}%
  \BibitemOpen
  \bibinfo {note} {See Supplementary Material for a detailed description, which
  includes references \cite
  {martinezpillet97,buehler15,rogers96,anders89,vogler04,beeck13,piskunov95,bellotrubio02,rutten82,shchukina01,krivova06}}\BibitemShut
  {NoStop}%
\bibitem [{\citenamefont {{V{\"o}gler}}\ \emph {et~al.}(2005)\citenamefont
  {{V{\"o}gler}}, \citenamefont {{Shelyag}}, \citenamefont {{Sch{\"u}ssler}},
  \citenamefont {{Cattaneo}}, \citenamefont {{Emonet}},\ and\ \citenamefont
  {{Linde}}}]{vogler05}%
  \BibitemOpen
  \bibfield  {author} {\bibinfo {author} {\bibfnamefont {A.}~\bibnamefont
  {{V{\"o}gler}}}, \bibinfo {author} {\bibfnamefont {S.}~\bibnamefont
  {{Shelyag}}}, \bibinfo {author} {\bibfnamefont {M.}~\bibnamefont
  {{Sch{\"u}ssler}}}, \bibinfo {author} {\bibfnamefont {F.}~\bibnamefont
  {{Cattaneo}}}, \bibinfo {author} {\bibfnamefont {T.}~\bibnamefont
  {{Emonet}}}, \ and\ \bibinfo {author} {\bibfnamefont {T.}~\bibnamefont
  {{Linde}}},\ }\href {\doibase 10.1051/0004-6361:20041507} {\bibfield
  {journal} {\bibinfo  {journal} {Astron. Astrophys.}\ }\textbf {\bibinfo
  {volume} {429}},\ \bibinfo {pages} {335} (\bibinfo {year}
  {2005})}\BibitemShut {NoStop}%
\bibitem [{\citenamefont {{Shelyag}}\ \emph {et~al.}(2004)\citenamefont
  {{Shelyag}}, \citenamefont {{Sch{\"u}ssler}}, \citenamefont {{Solanki}},
  \citenamefont {{Berdyugina}},\ and\ \citenamefont
  {{V{\"o}gler}}}]{shelyag04}%
  \BibitemOpen
  \bibfield  {author} {\bibinfo {author} {\bibfnamefont {S.}~\bibnamefont
  {{Shelyag}}}, \bibinfo {author} {\bibfnamefont {M.}~\bibnamefont
  {{Sch{\"u}ssler}}}, \bibinfo {author} {\bibfnamefont {S.~K.}\ \bibnamefont
  {{Solanki}}}, \bibinfo {author} {\bibfnamefont {S.~V.}\ \bibnamefont
  {{Berdyugina}}}, \ and\ \bibinfo {author} {\bibfnamefont {A.}~\bibnamefont
  {{V{\"o}gler}}},\ }\href {\doibase 10.1051/0004-6361:20040471} {\bibfield
  {journal} {\bibinfo  {journal} {Astron. Astrophys.}\ }\textbf {\bibinfo
  {volume} {427}},\ \bibinfo {pages} {335} (\bibinfo {year}
  {2004})}\BibitemShut {NoStop}%
\bibitem [{\citenamefont {{Afram}}\ \emph {et~al.}(2011)\citenamefont
  {{Afram}}, \citenamefont {{Unruh}}, \citenamefont {{Solanki}}, \citenamefont
  {{Sch{\"u}ssler}}, \citenamefont {{Lagg}},\ and\ \citenamefont
  {{V{\"o}gler}}}]{afram11}%
  \BibitemOpen
  \bibfield  {author} {\bibinfo {author} {\bibfnamefont {N.}~\bibnamefont
  {{Afram}}}, \bibinfo {author} {\bibfnamefont {Y.~C.}\ \bibnamefont
  {{Unruh}}}, \bibinfo {author} {\bibfnamefont {S.~K.}\ \bibnamefont
  {{Solanki}}}, \bibinfo {author} {\bibfnamefont {M.}~\bibnamefont
  {{Sch{\"u}ssler}}}, \bibinfo {author} {\bibfnamefont {A.}~\bibnamefont
  {{Lagg}}}, \ and\ \bibinfo {author} {\bibfnamefont {A.}~\bibnamefont
  {{V{\"o}gler}}},\ }\href {\doibase 10.1051/0004-6361/201015582} {\bibfield
  {journal} {\bibinfo  {journal} {Astron. Astrophys.}\ }\textbf {\bibinfo
  {volume} {526}},\ \bibinfo {pages} {A120} (\bibinfo {year}
  {2011})}\BibitemShut {NoStop}%
\bibitem [{\citenamefont {{Danilovic}}\ \emph {et~al.}(2013)\citenamefont
  {{Danilovic}}, \citenamefont {{R{\"o}hrbein}}, \citenamefont {{Cameron}},\
  and\ \citenamefont {{Sch{\"u}ssler}}}]{danilovic13}%
  \BibitemOpen
  \bibfield  {author} {\bibinfo {author} {\bibfnamefont {S.}~\bibnamefont
  {{Danilovic}}}, \bibinfo {author} {\bibfnamefont {D.}~\bibnamefont
  {{R{\"o}hrbein}}}, \bibinfo {author} {\bibfnamefont {R.~H.}\ \bibnamefont
  {{Cameron}}}, \ and\ \bibinfo {author} {\bibfnamefont {M.}~\bibnamefont
  {{Sch{\"u}ssler}}},\ }\href {\doibase 10.1051/0004-6361/201219726} {\bibfield
   {journal} {\bibinfo  {journal} {Astron. Astrophys.}\ }\textbf {\bibinfo
  {volume} {550}},\ \bibinfo {pages} {A118} (\bibinfo {year}
  {2013})}\BibitemShut {NoStop}%
\bibitem [{\citenamefont {{Riethm{\"u}ller}}\ \emph {et~al.}(2014)\citenamefont
  {{Riethm{\"u}ller}}, \citenamefont {{Solanki}}, \citenamefont {{Berdyugina}},
  \citenamefont {{Sch{\"u}ssler}}, \citenamefont {{Mart{\'{\i}}nez Pillet}},
  \citenamefont {{Feller}}, \citenamefont {{Gandorfer}},\ and\ \citenamefont
  {{Hirzberger}}}]{riethmuller14}%
  \BibitemOpen
  \bibfield  {author} {\bibinfo {author} {\bibfnamefont {T.~L.}\ \bibnamefont
  {{Riethm{\"u}ller}}}, \bibinfo {author} {\bibfnamefont {S.~K.}\ \bibnamefont
  {{Solanki}}}, \bibinfo {author} {\bibfnamefont {S.~V.}\ \bibnamefont
  {{Berdyugina}}}, \bibinfo {author} {\bibfnamefont {M.}~\bibnamefont
  {{Sch{\"u}ssler}}}, \bibinfo {author} {\bibfnamefont {V.}~\bibnamefont
  {{Mart{\'{\i}}nez Pillet}}}, \bibinfo {author} {\bibfnamefont
  {A.}~\bibnamefont {{Feller}}}, \bibinfo {author} {\bibfnamefont
  {A.}~\bibnamefont {{Gandorfer}}}, \ and\ \bibinfo {author} {\bibfnamefont
  {J.}~\bibnamefont {{Hirzberger}}},\ }\href {\doibase
  10.1051/0004-6361/201423892} {\bibfield  {journal} {\bibinfo  {journal}
  {Astron. Astrophys.}\ }\textbf {\bibinfo {volume} {568}},\ \bibinfo {pages}
  {A13} (\bibinfo {year} {2014})}\BibitemShut {NoStop}%
\bibitem [{\citenamefont {{Schou}}\ \emph {et~al.}(2012)\citenamefont
  {{Schou}}, \citenamefont {{Scherrer}}, \citenamefont {{Bush}}, \citenamefont
  {{Wachter}}, \citenamefont {{Couvidat}}, \citenamefont {{Rabello-Soares}},
  \citenamefont {{Bogart}}, \citenamefont {{Hoeksema}}, \citenamefont {{Liu}},
  \citenamefont {{Duvall}}, \citenamefont {{Akin}}, \citenamefont {{Allard}},
  \citenamefont {{Miles}}, \citenamefont {{Rairden}}, \citenamefont {{Shine}},
  \citenamefont {{Tarbell}}, \citenamefont {{Title}}, \citenamefont
  {{Wolfson}}, \citenamefont {{Elmore}}, \citenamefont {{Norton}},\ and\
  \citenamefont {{Tomczyk}}}]{schou12}%
  \BibitemOpen
  \bibfield  {author} {\bibinfo {author} {\bibfnamefont {J.}~\bibnamefont
  {{Schou}}}, \bibinfo {author} {\bibfnamefont {P.~H.}\ \bibnamefont
  {{Scherrer}}}, \bibinfo {author} {\bibfnamefont {R.~I.}\ \bibnamefont
  {{Bush}}}, \bibinfo {author} {\bibfnamefont {R.}~\bibnamefont {{Wachter}}},
  \bibinfo {author} {\bibfnamefont {S.}~\bibnamefont {{Couvidat}}}, \bibinfo
  {author} {\bibfnamefont {M.~C.}\ \bibnamefont {{Rabello-Soares}}}, \bibinfo
  {author} {\bibfnamefont {R.~S.}\ \bibnamefont {{Bogart}}}, \bibinfo {author}
  {\bibfnamefont {J.~T.}\ \bibnamefont {{Hoeksema}}}, \bibinfo {author}
  {\bibfnamefont {Y.}~\bibnamefont {{Liu}}}, \bibinfo {author} {\bibfnamefont
  {T.~L.}\ \bibnamefont {{Duvall}}}, \bibinfo {author} {\bibfnamefont {D.~J.}\
  \bibnamefont {{Akin}}}, \bibinfo {author} {\bibfnamefont {B.~A.}\
  \bibnamefont {{Allard}}}, \bibinfo {author} {\bibfnamefont {J.~W.}\
  \bibnamefont {{Miles}}}, \bibinfo {author} {\bibfnamefont {R.}~\bibnamefont
  {{Rairden}}}, \bibinfo {author} {\bibfnamefont {R.~A.}\ \bibnamefont
  {{Shine}}}, \bibinfo {author} {\bibfnamefont {T.~D.}\ \bibnamefont
  {{Tarbell}}}, \bibinfo {author} {\bibfnamefont {A.~M.}\ \bibnamefont
  {{Title}}}, \bibinfo {author} {\bibfnamefont {C.~J.}\ \bibnamefont
  {{Wolfson}}}, \bibinfo {author} {\bibfnamefont {D.~F.}\ \bibnamefont
  {{Elmore}}}, \bibinfo {author} {\bibfnamefont {A.~A.}\ \bibnamefont
  {{Norton}}}, \ and\ \bibinfo {author} {\bibfnamefont {S.}~\bibnamefont
  {{Tomczyk}}},\ }\href {\doibase 10.1007/s11207-011-9842-2} {\bibfield
  {journal} {\bibinfo  {journal} {Sol. Phys.}\ }\textbf {\bibinfo {volume}
  {275}},\ \bibinfo {pages} {229} (\bibinfo {year} {2012})}\BibitemShut
  {NoStop}%
\bibitem [{\citenamefont {{Yeo}}\ \emph
  {et~al.}(2014{\natexlab{b}})\citenamefont {{Yeo}}, \citenamefont {{Krivova}},
  \citenamefont {{Solanki}},\ and\ \citenamefont {{Glassmeier}}}]{yeo14a}%
  \BibitemOpen
  \bibfield  {author} {\bibinfo {author} {\bibfnamefont {K.~L.}\ \bibnamefont
  {{Yeo}}}, \bibinfo {author} {\bibfnamefont {N.~A.}\ \bibnamefont
  {{Krivova}}}, \bibinfo {author} {\bibfnamefont {S.~K.}\ \bibnamefont
  {{Solanki}}}, \ and\ \bibinfo {author} {\bibfnamefont {K.~H.}\ \bibnamefont
  {{Glassmeier}}},\ }\href {\doibase 10.1051/0004-6361/201423628} {\bibfield
  {journal} {\bibinfo  {journal} {Astron. Astrophys.}\ }\textbf {\bibinfo
  {volume} {570}},\ \bibinfo {pages} {A85} (\bibinfo {year}
  {2014}{\natexlab{b}})}\BibitemShut {NoStop}%
\bibitem [{\citenamefont {{Unruh}}\ \emph {et~al.}(1999)\citenamefont
  {{Unruh}}, \citenamefont {{Solanki}},\ and\ \citenamefont
  {{Fligge}}}]{unruh99}%
  \BibitemOpen
  \bibfield  {author} {\bibinfo {author} {\bibfnamefont {Y.~C.}\ \bibnamefont
  {{Unruh}}}, \bibinfo {author} {\bibfnamefont {S.~K.}\ \bibnamefont
  {{Solanki}}}, \ and\ \bibinfo {author} {\bibfnamefont {M.}~\bibnamefont
  {{Fligge}}},\ }\href@noop {} {\bibfield  {journal} {\bibinfo  {journal}
  {Astron. Astrophys.}\ }\textbf {\bibinfo {volume} {345}},\ \bibinfo {pages}
  {635} (\bibinfo {year} {1999})}\BibitemShut {NoStop}%
\bibitem [{\citenamefont {{Title}}\ \emph {et~al.}(1992)\citenamefont
  {{Title}}, \citenamefont {{Topka}}, \citenamefont {{Tarbell}}, \citenamefont
  {{Schmidt}}, \citenamefont {{Balke}},\ and\ \citenamefont
  {{Scharmer}}}]{title92}%
  \BibitemOpen
  \bibfield  {author} {\bibinfo {author} {\bibfnamefont {A.~M.}\ \bibnamefont
  {{Title}}}, \bibinfo {author} {\bibfnamefont {K.~P.}\ \bibnamefont
  {{Topka}}}, \bibinfo {author} {\bibfnamefont {T.~D.}\ \bibnamefont
  {{Tarbell}}}, \bibinfo {author} {\bibfnamefont {W.}~\bibnamefont
  {{Schmidt}}}, \bibinfo {author} {\bibfnamefont {C.}~\bibnamefont {{Balke}}},
  \ and\ \bibinfo {author} {\bibfnamefont {G.}~\bibnamefont {{Scharmer}}},\
  }\href {\doibase 10.1086/171545} {\bibfield  {journal} {\bibinfo  {journal}
  {Astrophys. J.}\ }\textbf {\bibinfo {volume} {393}},\ \bibinfo {pages} {782}
  (\bibinfo {year} {1992})}\BibitemShut {NoStop}%
\bibitem [{\citenamefont {{Ortiz}}\ \emph {et~al.}(2002)\citenamefont
  {{Ortiz}}, \citenamefont {{Solanki}}, \citenamefont {{Domingo}},
  \citenamefont {{Fligge}},\ and\ \citenamefont {{Sanahuja}}}]{ortiz02}%
  \BibitemOpen
  \bibfield  {author} {\bibinfo {author} {\bibfnamefont {A.}~\bibnamefont
  {{Ortiz}}}, \bibinfo {author} {\bibfnamefont {S.~K.}\ \bibnamefont
  {{Solanki}}}, \bibinfo {author} {\bibfnamefont {V.}~\bibnamefont
  {{Domingo}}}, \bibinfo {author} {\bibfnamefont {M.}~\bibnamefont {{Fligge}}},
  \ and\ \bibinfo {author} {\bibfnamefont {B.}~\bibnamefont {{Sanahuja}}},\
  }\href {\doibase 10.1051/0004-6361:20020500} {\bibfield  {journal} {\bibinfo
  {journal} {Astron. Astrophys.}\ }\textbf {\bibinfo {volume} {388}},\ \bibinfo
  {pages} {1036} (\bibinfo {year} {2002})}\BibitemShut {NoStop}%
\bibitem [{\citenamefont {{Kurucz}}(1992)}]{kurucz92}%
  \BibitemOpen
  \bibfield  {author} {\bibinfo {author} {\bibfnamefont {R.~L.}\ \bibnamefont
  {{Kurucz}}},\ }\href@noop {} {\bibfield  {journal} {\bibinfo  {journal} {Rev.
  Mex. Astron. Astrofis.}\ }\textbf {\bibinfo {volume} {23}} (\bibinfo {year}
  {1992})}\BibitemShut {NoStop}%
\bibitem [{\citenamefont {{Frutiger}}\ \emph {et~al.}(2000)\citenamefont
  {{Frutiger}}, \citenamefont {{Solanki}}, \citenamefont {{Fligge}},\ and\
  \citenamefont {{Bruls}}}]{frutiger00}%
  \BibitemOpen
  \bibfield  {author} {\bibinfo {author} {\bibfnamefont {C.}~\bibnamefont
  {{Frutiger}}}, \bibinfo {author} {\bibfnamefont {S.~K.}\ \bibnamefont
  {{Solanki}}}, \bibinfo {author} {\bibfnamefont {M.}~\bibnamefont {{Fligge}}},
  \ and\ \bibinfo {author} {\bibfnamefont {J.~H.~M.~J.}\ \bibnamefont
  {{Bruls}}},\ }\href@noop {} {\bibfield  {journal} {\bibinfo  {journal}
  {Astron. Astrophys.}\ }\textbf {\bibinfo {volume} {358}},\ \bibinfo {pages}
  {1109} (\bibinfo {year} {2000})}\BibitemShut {NoStop}%
\bibitem [{\citenamefont {{Yeo}}\ \emph
  {et~al.}(2014{\natexlab{c}})\citenamefont {{Yeo}}, \citenamefont {{Feller}},
  \citenamefont {{Solanki}}, \citenamefont {{Couvidat}}, \citenamefont
  {{Danilovic}},\ and\ \citenamefont {{Krivova}}}]{yeo14c}%
  \BibitemOpen
  \bibfield  {author} {\bibinfo {author} {\bibfnamefont {K.~L.}\ \bibnamefont
  {{Yeo}}}, \bibinfo {author} {\bibfnamefont {A.}~\bibnamefont {{Feller}}},
  \bibinfo {author} {\bibfnamefont {S.~K.}\ \bibnamefont {{Solanki}}}, \bibinfo
  {author} {\bibfnamefont {S.}~\bibnamefont {{Couvidat}}}, \bibinfo {author}
  {\bibfnamefont {S.}~\bibnamefont {{Danilovic}}}, \ and\ \bibinfo {author}
  {\bibfnamefont {N.~A.}\ \bibnamefont {{Krivova}}},\ }\href {\doibase
  10.1051/0004-6361/201322502} {\bibfield  {journal} {\bibinfo  {journal}
  {Astron. Astrophys.}\ }\textbf {\bibinfo {volume} {561}},\ \bibinfo {pages}
  {A22} (\bibinfo {year} {2014}{\natexlab{c}})}\BibitemShut {NoStop}%
\bibitem [{\citenamefont {{Couvidat}}\ \emph {et~al.}(2012)\citenamefont
  {{Couvidat}}, \citenamefont {{Rajaguru}}, \citenamefont {{Wachter}},
  \citenamefont {{Sankarasubramanian}}, \citenamefont {{Schou}},\ and\
  \citenamefont {{Scherrer}}}]{couvidat12}%
  \BibitemOpen
  \bibfield  {author} {\bibinfo {author} {\bibfnamefont {S.}~\bibnamefont
  {{Couvidat}}}, \bibinfo {author} {\bibfnamefont {S.~P.}\ \bibnamefont
  {{Rajaguru}}}, \bibinfo {author} {\bibfnamefont {R.}~\bibnamefont
  {{Wachter}}}, \bibinfo {author} {\bibfnamefont {K.}~\bibnamefont
  {{Sankarasubramanian}}}, \bibinfo {author} {\bibfnamefont {J.}~\bibnamefont
  {{Schou}}}, \ and\ \bibinfo {author} {\bibfnamefont {P.~H.}\ \bibnamefont
  {{Scherrer}}},\ }\href {\doibase 10.1007/s11207-011-9927-y} {\bibfield
  {journal} {\bibinfo  {journal} {Sol. Phys.}\ }\textbf {\bibinfo {volume}
  {278}},\ \bibinfo {pages} {217} (\bibinfo {year} {2012})}\BibitemShut
  {NoStop}%
\bibitem [{\citenamefont {{Kopp}}\ \emph {et~al.}(2005)\citenamefont {{Kopp}},
  \citenamefont {{Lawrence}},\ and\ \citenamefont {{Rottman}}}]{kopp05}%
  \BibitemOpen
  \bibfield  {author} {\bibinfo {author} {\bibfnamefont {G.}~\bibnamefont
  {{Kopp}}}, \bibinfo {author} {\bibfnamefont {G.}~\bibnamefont {{Lawrence}}},
  \ and\ \bibinfo {author} {\bibfnamefont {G.}~\bibnamefont {{Rottman}}},\
  }\href {\doibase 10.1007/s11207-005-7433-9} {\bibfield  {journal} {\bibinfo
  {journal} {Sol. Phys.}\ }\textbf {\bibinfo {volume} {230}},\ \bibinfo {pages}
  {129} (\bibinfo {year} {2005})}\BibitemShut {NoStop}%
\bibitem [{\citenamefont {{Fr{\"o}hlich}}\ \emph {et~al.}(1995)\citenamefont
  {{Fr{\"o}hlich}}, \citenamefont {{Romero}}, \citenamefont {{Roth}},
  \citenamefont {{Wehrli}}, \citenamefont {{Andersen}}, \citenamefont
  {{Appourchaux}}, \citenamefont {{Domingo}}, \citenamefont {{Telljohann}},
  \citenamefont {{Berthomieu}}, \citenamefont {{Delache}}, \citenamefont
  {{Provost}}, \citenamefont {{Toutain}}, \citenamefont {{Crommelynck}},
  \citenamefont {{Chevalier}}, \citenamefont {{Fichot}}, \citenamefont
  {{D{\"a}ppen}}, \citenamefont {{Gough}}, \citenamefont {{Hoeksema}},
  \citenamefont {{Jim{\'e}nez}}, \citenamefont {{G{\'o}mez}}, \citenamefont
  {{Herreros}}, \citenamefont {{Cort{\'e}s}}, \citenamefont {{Jones}},
  \citenamefont {{Pap}},\ and\ \citenamefont {{Willson}}}]{frohlich95}%
  \BibitemOpen
  \bibfield  {author} {\bibinfo {author} {\bibfnamefont {C.}~\bibnamefont
  {{Fr{\"o}hlich}}}, \bibinfo {author} {\bibfnamefont {J.}~\bibnamefont
  {{Romero}}}, \bibinfo {author} {\bibfnamefont {H.}~\bibnamefont {{Roth}}},
  \bibinfo {author} {\bibfnamefont {C.}~\bibnamefont {{Wehrli}}}, \bibinfo
  {author} {\bibfnamefont {B.~N.}\ \bibnamefont {{Andersen}}}, \bibinfo
  {author} {\bibfnamefont {T.}~\bibnamefont {{Appourchaux}}}, \bibinfo {author}
  {\bibfnamefont {V.}~\bibnamefont {{Domingo}}}, \bibinfo {author}
  {\bibfnamefont {U.}~\bibnamefont {{Telljohann}}}, \bibinfo {author}
  {\bibfnamefont {G.}~\bibnamefont {{Berthomieu}}}, \bibinfo {author}
  {\bibfnamefont {P.}~\bibnamefont {{Delache}}}, \bibinfo {author}
  {\bibfnamefont {J.}~\bibnamefont {{Provost}}}, \bibinfo {author}
  {\bibfnamefont {T.}~\bibnamefont {{Toutain}}}, \bibinfo {author}
  {\bibfnamefont {D.~A.}\ \bibnamefont {{Crommelynck}}}, \bibinfo {author}
  {\bibfnamefont {A.}~\bibnamefont {{Chevalier}}}, \bibinfo {author}
  {\bibfnamefont {A.}~\bibnamefont {{Fichot}}}, \bibinfo {author}
  {\bibfnamefont {W.}~\bibnamefont {{D{\"a}ppen}}}, \bibinfo {author}
  {\bibfnamefont {D.}~\bibnamefont {{Gough}}}, \bibinfo {author} {\bibfnamefont
  {T.}~\bibnamefont {{Hoeksema}}}, \bibinfo {author} {\bibfnamefont
  {A.}~\bibnamefont {{Jim{\'e}nez}}}, \bibinfo {author} {\bibfnamefont {M.~F.}\
  \bibnamefont {{G{\'o}mez}}}, \bibinfo {author} {\bibfnamefont {J.~M.}\
  \bibnamefont {{Herreros}}}, \bibinfo {author} {\bibfnamefont {T.~R.}\
  \bibnamefont {{Cort{\'e}s}}}, \bibinfo {author} {\bibfnamefont {A.~R.}\
  \bibnamefont {{Jones}}}, \bibinfo {author} {\bibfnamefont {J.~M.}\
  \bibnamefont {{Pap}}}, \ and\ \bibinfo {author} {\bibfnamefont {R.~C.}\
  \bibnamefont {{Willson}}},\ }\href {\doibase 10.1007/BF00733428} {\bibfield
  {journal} {\bibinfo  {journal} {Sol. Phys.}\ }\textbf {\bibinfo {volume}
  {162}},\ \bibinfo {pages} {101} (\bibinfo {year} {1995})}\BibitemShut
  {NoStop}%
\bibitem [{\citenamefont {{Kopp}}\ and\ \citenamefont {{Lean}}(2011)}]{kopp11}%
  \BibitemOpen
  \bibfield  {author} {\bibinfo {author} {\bibfnamefont {G.}~\bibnamefont
  {{Kopp}}}\ and\ \bibinfo {author} {\bibfnamefont {J.~L.}\ \bibnamefont
  {{Lean}}},\ }\href {\doibase 10.1029/2010GL045777} {\bibfield  {journal}
  {\bibinfo  {journal} {Geophys. Res. Lett.}\ }\textbf {\bibinfo {volume}
  {38}},\ \bibinfo {pages} {1706} (\bibinfo {year} {2011})}\BibitemShut
  {NoStop}%
\bibitem [{\citenamefont {{Kopp}}\ \emph {et~al.}(2012)\citenamefont {{Kopp}},
  \citenamefont {{Fehlmann}}, \citenamefont {{Finsterle}}, \citenamefont
  {{Harber}}, \citenamefont {{Heuerman}},\ and\ \citenamefont
  {{Willson}}}]{kopp12}%
  \BibitemOpen
  \bibfield  {author} {\bibinfo {author} {\bibfnamefont {G.}~\bibnamefont
  {{Kopp}}}, \bibinfo {author} {\bibfnamefont {A.}~\bibnamefont {{Fehlmann}}},
  \bibinfo {author} {\bibfnamefont {W.}~\bibnamefont {{Finsterle}}}, \bibinfo
  {author} {\bibfnamefont {D.}~\bibnamefont {{Harber}}}, \bibinfo {author}
  {\bibfnamefont {K.}~\bibnamefont {{Heuerman}}}, \ and\ \bibinfo {author}
  {\bibfnamefont {R.}~\bibnamefont {{Willson}}},\ }\href {\doibase
  10.1088/0026-1394/49/2/S29} {\bibfield  {journal} {\bibinfo  {journal}
  {Metrologia}\ }\textbf {\bibinfo {volume} {49}},\ \bibinfo {pages} {29}
  (\bibinfo {year} {2012})}\BibitemShut {NoStop}%
\bibitem [{\citenamefont {{Fehlmann}}\ \emph {et~al.}(2012)\citenamefont
  {{Fehlmann}}, \citenamefont {{Kopp}}, \citenamefont {{Schmutz}},
  \citenamefont {{Winkler}}, \citenamefont {{Finsterle}},\ and\ \citenamefont
  {{Fox}}}]{fehlmann12}%
  \BibitemOpen
  \bibfield  {author} {\bibinfo {author} {\bibfnamefont {A.}~\bibnamefont
  {{Fehlmann}}}, \bibinfo {author} {\bibfnamefont {G.}~\bibnamefont {{Kopp}}},
  \bibinfo {author} {\bibfnamefont {W.}~\bibnamefont {{Schmutz}}}, \bibinfo
  {author} {\bibfnamefont {R.}~\bibnamefont {{Winkler}}}, \bibinfo {author}
  {\bibfnamefont {W.}~\bibnamefont {{Finsterle}}}, \ and\ \bibinfo {author}
  {\bibfnamefont {N.}~\bibnamefont {{Fox}}},\ }\href {\doibase
  10.1088/0026-1394/49/2/S34} {\bibfield  {journal} {\bibinfo  {journal}
  {Metrologia}\ }\textbf {\bibinfo {volume} {49}},\ \bibinfo {pages} {34}
  (\bibinfo {year} {2012})}\BibitemShut {NoStop}%
\bibitem [{\citenamefont {{Mart{\'{\i}}nez Pillet}}\ \emph
  {et~al.}(1997)\citenamefont {{Mart{\'{\i}}nez Pillet}}, \citenamefont
  {{Lites}},\ and\ \citenamefont {{Skumanich}}}]{martinezpillet97}%
  \BibitemOpen
  \bibfield  {author} {\bibinfo {author} {\bibfnamefont {V.}~\bibnamefont
  {{Mart{\'{\i}}nez Pillet}}}, \bibinfo {author} {\bibfnamefont {B.~W.}\
  \bibnamefont {{Lites}}}, \ and\ \bibinfo {author} {\bibfnamefont
  {A.}~\bibnamefont {{Skumanich}}},\ }\href {\doibase 10.1086/303478}
  {\bibfield  {journal} {\bibinfo  {journal} {Astrophys. J.}\ }\textbf
  {\bibinfo {volume} {474}},\ \bibinfo {pages} {810} (\bibinfo {year}
  {1997})}\BibitemShut {NoStop}%
\bibitem [{\citenamefont {{Buehler}}\ \emph {et~al.}(2015)\citenamefont
  {{Buehler}}, \citenamefont {{Lagg}}, \citenamefont {{Solanki}},\ and\
  \citenamefont {{van Noort}}}]{buehler15}%
  \BibitemOpen
  \bibfield  {author} {\bibinfo {author} {\bibfnamefont {D.}~\bibnamefont
  {{Buehler}}}, \bibinfo {author} {\bibfnamefont {A.}~\bibnamefont {{Lagg}}},
  \bibinfo {author} {\bibfnamefont {S.~K.}\ \bibnamefont {{Solanki}}}, \ and\
  \bibinfo {author} {\bibfnamefont {M.}~\bibnamefont {{van Noort}}},\ }\href
  {\doibase 10.1051/0004-6361/201424970} {\bibfield  {journal} {\bibinfo
  {journal} {Astron. Astrophys.}\ }\textbf {\bibinfo {volume} {576}},\ \bibinfo
  {pages} {A27} (\bibinfo {year} {2015})}\BibitemShut {NoStop}%
\bibitem [{\citenamefont {{Rogers}}\ \emph {et~al.}(1996)\citenamefont
  {{Rogers}}, \citenamefont {{Swenson}},\ and\ \citenamefont
  {{Iglesias}}}]{rogers96}%
  \BibitemOpen
  \bibfield  {author} {\bibinfo {author} {\bibfnamefont {F.~J.}\ \bibnamefont
  {{Rogers}}}, \bibinfo {author} {\bibfnamefont {F.~J.}\ \bibnamefont
  {{Swenson}}}, \ and\ \bibinfo {author} {\bibfnamefont {C.~A.}\ \bibnamefont
  {{Iglesias}}},\ }\href {\doibase 10.1086/176705} {\bibfield  {journal}
  {\bibinfo  {journal} {Astrophys. J.}\ }\textbf {\bibinfo {volume} {456}},\
  \bibinfo {pages} {902} (\bibinfo {year} {1996})}\BibitemShut {NoStop}%
\bibitem [{\citenamefont {{Anders}}\ and\ \citenamefont
  {{Grevesse}}(1989)}]{anders89}%
  \BibitemOpen
  \bibfield  {author} {\bibinfo {author} {\bibfnamefont {E.}~\bibnamefont
  {{Anders}}}\ and\ \bibinfo {author} {\bibfnamefont {N.}~\bibnamefont
  {{Grevesse}}},\ }\href {\doibase 10.1016/0016-7037(89)90286-X} {\bibfield
  {journal} {\bibinfo  {journal} {Geochim. Cosmochim. Acta.}\ }\textbf
  {\bibinfo {volume} {53}},\ \bibinfo {pages} {197} (\bibinfo {year}
  {1989})}\BibitemShut {NoStop}%
\bibitem [{\citenamefont {{V{\"o}gler}}\ \emph {et~al.}(2004)\citenamefont
  {{V{\"o}gler}}, \citenamefont {{Bruls}},\ and\ \citenamefont
  {{Sch{\"u}ssler}}}]{vogler04}%
  \BibitemOpen
  \bibfield  {author} {\bibinfo {author} {\bibfnamefont {A.}~\bibnamefont
  {{V{\"o}gler}}}, \bibinfo {author} {\bibfnamefont {J.~H.~M.~J.}\ \bibnamefont
  {{Bruls}}}, \ and\ \bibinfo {author} {\bibfnamefont {M.}~\bibnamefont
  {{Sch{\"u}ssler}}},\ }\href {\doibase 10.1051/0004-6361:20047043} {\bibfield
  {journal} {\bibinfo  {journal} {Astron. Astrophys.}\ }\textbf {\bibinfo
  {volume} {421}},\ \bibinfo {pages} {741} (\bibinfo {year}
  {2004})}\BibitemShut {NoStop}%
\bibitem [{\citenamefont {{Beeck}}\ \emph {et~al.}(2013)\citenamefont
  {{Beeck}}, \citenamefont {{Cameron}}, \citenamefont {{Reiners}},\ and\
  \citenamefont {{Sch{\"u}ssler}}}]{beeck13}%
  \BibitemOpen
  \bibfield  {author} {\bibinfo {author} {\bibfnamefont {B.}~\bibnamefont
  {{Beeck}}}, \bibinfo {author} {\bibfnamefont {R.~H.}\ \bibnamefont
  {{Cameron}}}, \bibinfo {author} {\bibfnamefont {A.}~\bibnamefont
  {{Reiners}}}, \ and\ \bibinfo {author} {\bibfnamefont {M.}~\bibnamefont
  {{Sch{\"u}ssler}}},\ }\href {\doibase 10.1051/0004-6361/201321343} {\bibfield
   {journal} {\bibinfo  {journal} {Astron. Astrophys.}\ }\textbf {\bibinfo
  {volume} {558}},\ \bibinfo {pages} {A48} (\bibinfo {year}
  {2013})}\BibitemShut {NoStop}%
\bibitem [{\citenamefont {{Piskunov}}\ \emph {et~al.}(1995)\citenamefont
  {{Piskunov}}, \citenamefont {{Kupka}}, \citenamefont {{Ryabchikova}},
  \citenamefont {{Weiss}},\ and\ \citenamefont {{Jeffery}}}]{piskunov95}%
  \BibitemOpen
  \bibfield  {author} {\bibinfo {author} {\bibfnamefont {N.~E.}\ \bibnamefont
  {{Piskunov}}}, \bibinfo {author} {\bibfnamefont {F.}~\bibnamefont {{Kupka}}},
  \bibinfo {author} {\bibfnamefont {T.~A.}\ \bibnamefont {{Ryabchikova}}},
  \bibinfo {author} {\bibfnamefont {W.~W.}\ \bibnamefont {{Weiss}}}, \ and\
  \bibinfo {author} {\bibfnamefont {C.~S.}\ \bibnamefont {{Jeffery}}},\
  }\href@noop {} {\bibfield  {journal} {\bibinfo  {journal} {Astron. Astrophys.
  Suppl. Ser.}\ }\textbf {\bibinfo {volume} {112}},\ \bibinfo {pages} {525}
  (\bibinfo {year} {1995})}\BibitemShut {NoStop}%
\bibitem [{\citenamefont {{Bellot Rubio}}\ and\ \citenamefont
  {{Borrero}}(2002)}]{bellotrubio02}%
  \BibitemOpen
  \bibfield  {author} {\bibinfo {author} {\bibfnamefont {L.~R.}\ \bibnamefont
  {{Bellot Rubio}}}\ and\ \bibinfo {author} {\bibfnamefont {J.~M.}\
  \bibnamefont {{Borrero}}},\ }\href {\doibase 10.1051/0004-6361:20020656}
  {\bibfield  {journal} {\bibinfo  {journal} {Astron. Astrophys.}\ }\textbf
  {\bibinfo {volume} {391}},\ \bibinfo {pages} {331} (\bibinfo {year}
  {2002})}\BibitemShut {NoStop}%
\bibitem [{\citenamefont {{Rutten}}\ and\ \citenamefont
  {{Kostik}}(1982)}]{rutten82}%
  \BibitemOpen
  \bibfield  {author} {\bibinfo {author} {\bibfnamefont {R.~J.}\ \bibnamefont
  {{Rutten}}}\ and\ \bibinfo {author} {\bibfnamefont {R.~I.}\ \bibnamefont
  {{Kostik}}},\ }\href@noop {} {\bibfield  {journal} {\bibinfo  {journal}
  {Astron. Astrophys.}\ }\textbf {\bibinfo {volume} {115}},\ \bibinfo {pages}
  {104} (\bibinfo {year} {1982})}\BibitemShut {NoStop}%
\bibitem [{\citenamefont {{Shchukina}}\ and\ \citenamefont {{Trujillo
  Bueno}}(2001)}]{shchukina01}%
  \BibitemOpen
  \bibfield  {author} {\bibinfo {author} {\bibfnamefont {N.}~\bibnamefont
  {{Shchukina}}}\ and\ \bibinfo {author} {\bibfnamefont {J.}~\bibnamefont
  {{Trujillo Bueno}}},\ }\href {\doibase 10.1086/319789} {\bibfield  {journal}
  {\bibinfo  {journal} {Astrophys. J.}\ }\textbf {\bibinfo {volume} {550}},\
  \bibinfo {pages} {970} (\bibinfo {year} {2001})}\BibitemShut {NoStop}%
\bibitem [{\citenamefont {{Krivova}}\ \emph {et~al.}(2006)\citenamefont
  {{Krivova}}, \citenamefont {{Solanki}},\ and\ \citenamefont
  {{Floyd}}}]{krivova06}%
  \BibitemOpen
  \bibfield  {author} {\bibinfo {author} {\bibfnamefont {N.~A.}\ \bibnamefont
  {{Krivova}}}, \bibinfo {author} {\bibfnamefont {S.~K.}\ \bibnamefont
  {{Solanki}}}, \ and\ \bibinfo {author} {\bibfnamefont {L.}~\bibnamefont
  {{Floyd}}},\ }\href {\doibase 10.1051/0004-6361:20064809} {\bibfield
  {journal} {\bibinfo  {journal} {Astron. Astrophys.}\ }\textbf {\bibinfo
  {volume} {452}},\ \bibinfo {pages} {631} (\bibinfo {year}
  {2006})}\BibitemShut {NoStop}%
\end{thebibliography}

%

\end{document}